\documentclass[fleqn,usenatbib]{mnras}
\usepackage{newtxtext,newtxmath}
\usepackage[T1]{fontenc}

\DeclareRobustCommand{\VAN}[3]{#2}
\let\VANthebibliography\thebibliography
\def\thebibliography{\DeclareRobustCommand{\VAN}[3]{##3}\VANthebibliography}

\usepackage{xspace}
\usepackage{tcolorbox}

\newcommand{\FeKa}{Fe K$\alpha$ }
\newcommand{\software}[1]{{\sc #1}}
\newcommand{\Gradus}{\software{Gradus.jl}\xspace}
\newcommand{\relline}{\software{relline}\xspace}

\newcommand{\e}{\text{e}}
\renewcommand{\d}{\text{d}}
\newcommand{\rg}{r_\text{g}}
\newcommand{\utensor}[3]{#1^{#2}_{\phantom{#2}#3}}
\newcommand{\dtensor}[3]{#1_{#2}^{\phantom{#2}#3}}

\newcommand{\pderiv}[2]{\frac{\partial #1}{\partial #2}}
\newcommand{\risco}{r_\text{ISCO}}
\newcommand{\rhoem}{r_\text{em}}

\newcommand{\vel}[1]{v^{#1}}
\renewcommand{\vector}[1]{\bmath{#1}}
\newcommand{\jacobian}[2]{\left\lvert \frac{\partial #1}{\partial #2} \right\rvert}

\renewcommand{\Im}[1]{\text{Im}\left[#1\right]}
\renewcommand{\Re}[1]{\text{Re}\left[#1\right]}

\title[Gradus.jl]{Gradus.jl: spacetime-agnostic general relativistic ray-tracing
for X-ray spectral modelling}
\author[F. J. E. Baker et al.]{
F. J. E. Baker,$^{1}$\thanks{E-mail: fergus.baker@bristol.ac.uk (FB)}
and A. J. Young$^{1}$
\\
$^{1}$H. H. Wills Physics Laboratory, Tyndall Avenue, Bristol BS8 1TL, UK
}
\date{Accepted XXX. Received YYY; in original form ZZZ}
\pubyear{2025}

\begin{document}
\label{firstpage}
\pagerange{\pageref{firstpage}--\pageref{lastpage}}
\maketitle

\begin{abstract}
We introduce \Gradus, an open-source and publicly available general relativistic ray-tracing toolkit for spectral modelling in arbitrary spacetimes. Our software is written in the Julia programming language, making use of forward-mode automatic differentiation for computing the Christoffel symbols during geodesic integration, and for propagating derivatives through the entire ray-tracer. Relevant numerical methods are detailed, and our models are validated using a number of tests and comparisons to other codes. The differentiability is used to optimally calculate Cunningham transfer functions -- used to efficiently pre-compute relativistic effects in spectral models. A method is described for calculating such transfer functions for disc with non-zero vertical height, including the treatment of self-obscuration. An extension of the transfer function formalism that includes timing information is described, and used to calculate high-resolution reverberation lag spectra for a lamppost corona. The lag--frequency and lag--energy spectra for a Shakura--Sunyaev accretion disc with various lamppost heights and Eddington ratios are calculated, and the general impact of disc thickness in reflection models is discussed.
\end{abstract}

%Don't make up new ones.

%%%%%%%%%%%%%%%%%%%%%%%%%%%%%%%%%%%%%%%%%%%%%%%%%%

%%%%%%%%%%%%%%%%% BODY OF PAPER %%%%%%%%%%%%%%%%%%

%%% INTRODUCTION %%%%%%%%%%%%%%%%%%%%%%%%%%%%%%%%%
\section{Introduction}

General relativistic ray-tracing (GRRT) is a computational technique used to
calculate the trajectory of individual particles through a spacetime.  It
enables the simulation of photons and radiative processes in the
strong gravity around black holes, neutron stars, or other compact objects, and
is therefore invaluable for models of the inner regions of the accretion flow.
In this region, the general relativistic (GR) effects cause significant
deviations from the classical results in the observed spectra
\citep[e.g.][]{cunningham_optical_1973, fabian_long_2002}, timing
\citep[e.g.][]{stella_measuring_1990, reynolds_x-ray_1999}, and appearance
\citep[e.g.][]{luminet_image_1979}.

GRRT has a long history in spectral modelling, with widespread use in
calculating coronal illumination and accretion disc emissivity
\citep[e.g.][]{wilkins_understanding_2012, wilkins_towards_2016}, reflection
spectra \citep{fabian_x-ray_1989}, emission and absorption features
\citep[e.g.][]{ruszkowski_absorption_2002}, spectral energy densities,
\citep[e.g.][]{hagen_estimating_2023}, variabilities such as reverberation lags
\citep[e.g.][]{ingram_public_2019}, the disc-corona
connection in thermal lags \citep[e.g.][]{kammoun_hard_2019}, quasi-periodic
oscillations \citep[QPOs, e.g.][]{tsang_iron_2013}, and so on. We refer the
reader to the reviews of \citet{reynolds_iron_lines_2003} and
\citet{reynolds_observational_2021} for detailed discussion of relativistic
X-ray spectroscopy, and \citet{uttley_x-ray_2014} and
\citet{cackett_reverberation_2021} for reviews of X-ray reverberation, and the
accompanying use of GRRT.

Spectral models computed using GRRT commonly pre-compute the relativistic
effects, so that the models may be rapidly evaluated without invoking the
ray-tracer, thereby opening the opportunity for parameter inference using
optimisation methods. For example, in reflection spectra models, such as
\software{relline} / \software{relxill} \citep{dauser_broad_2010,
dauser_relativistic_2016} or \software{kyn} \citep{dovciak_extended_2004},
pre-computed tabular GRRT data is integrated. A consequence of pre-computation
is that only a select parameter domain is available and at limited resolution,
and assumptions concerning the disc geometry or velocity structure may become
baked into the tabulated data.  This may be problematic, as many GRRT-based
models assume, for computational simplicity, a razor-thin disc in the equatorial
plane with strictly Keplerian velocity structure
\citep[e.g.][]{laor_line_1991,dovciak_extended_2004, beckwith_iron_2004,
brenneman_constraining_2006, dauser_broad_2010}. Another model used in tandem
may make different assumptions, resulting in an inconsistent model. There is
therefore an interest in publicly available and performant GRRT codes that can
be used to account for and pre-compute GR effects, and can be made to
match the assumptions of other models.

In recent years, there is increased interest in `tests of relativity', studying
spacetime solutions that deviate from a Kerr black hole (e.g.
\citealt{johannsen_testing_2010, chrusciel_stationary_2012, bambi_testing_2022,
patra_accretion_2023, chen_observational_2024},). A line of research looks to
examine X-ray spectral properties of spacetimes that violate the
\textit{no-hair} uniqueness theorems\footnote{In \emph{no-hair} theorems, `hair'
    is metaphor for additional information in the spacetime beyond mass, angular
momentum, and charge.}. This recent interest is in part driven by observations
made by the Event Horizon Telescope collaboration of the `shadow' and photon
ring of M87 and Sgr A* \citep{the_event_horizon_telescope_collaboration_first_2019,
the_event_horizon_telescope_collaboration_first_2023}, as these have inspired a
wealth of study into different spacetimes and the development of new GRRT models
that are not tied to a particular geometry
\citep[see e.g.][]{eht_non_kerr_2022}. Other efforts use disc spectra and line
profiles to measure deviations from the Kerr spacetime, yielding seemingly tight
constraints on the deformation parameters of some spacetime solutions
\citep[e.g.][]{bambi_precision_measuremets_2021}. To explore the use and
validity of X-ray spectroscopy for such tests of relativity requires flexible
GRRT codes that are not specialised to a particular geometry.

For many spectral modelling applications, the pre-computed tabular GRRT data is a
set of \textit{transfer functions} that can be used to efficiently include
GR effects. There are few immediately available public codes that can reliably
calculate new transfer function tables, even assuming razor-thin Keplerian
disc models, and none that the authors are aware of that can reliably do so for
other disc geometries, velocity structures, and spacetimes. Perhaps the closest
example is the work of \citep{taylor_exploring_2018}, however this has some
caveats related to calculating transfer function tables. It is to address this
gap that we present a new publicly available GRRT code,
\Gradus\footnote{Available under GPL 3.0 license at
\url{https://github.com/astro-group-bristol/Gradus.jl}. Please submit bug
reports either under Issues, or directly via email to the corresponding
author.}, written in the Julia programming language
\citep{Bezanson_Julia_A_fresh_2017}.

The paper is organised in the following manner: in
Section~\ref{sec:numerical-methods} we describe the numerical methods used in
\Gradus, introducing the techniques and algorithms of our code. This description
necessarily avoids some implementation details, focusing on the mathematical and
algorithmic narrative, as the implementation is publicly available and
meticulously described in the \Gradus documentation. In
Section~\ref{sec:description-of-code} the software is detailed in terms of its
features. Section~\ref{sec:test-problems} assesses the validity of the code
using a number of test problems from the literature, and compares results to
other published codes.  Section~\ref{sec:applications} shows illustrative
simulation results that examine the effect of disc thickness on the line profile
and reverberation lag computed using \Gradus. Some conclusions, intended
applications, and future work are given in Section~\ref{sec:conclusion}.

%%% NUMERICAL METHODS %%%%%%%%%%%%%%%%%%%%%%%%%%%%
\section{Numerical methods}
\label{sec:numerical-methods}

We focus on stationary, axisymmetric, and asymptotically flat spacetimes in the
Boyer--Lindquist coordinates. Such spacetimes have metrics of the form
\begin{equation}
\label{eq:stationary_axisymmetric_metric}
    g_{\mu\nu}
    = g_{tt} \d t^2
    + g_{rr} \d r^2
    + g_{\theta\theta} \d \theta^2
    + g_{\phi\phi} \d \phi^2
    + 2g_{t\phi} \d t \d \phi.
\end{equation}
We adopt $(-, +, +, +)$ metric signature, and standard units $c = G
= 1$. Greek indices ($\mu, \nu$) denote the four spacetime components, and Latin
indices ($i, j$) denote the three spatial components. We write partial
derivatives with respect to the coordinates $x^\mu$ as $\partial_\mu := \partial
/ \partial x^\mu$.

\subsection{Geodesic integration}

The trajectory of particles in curved space is determined by integrating the
geodesic equation, termed \emph{ray-tracing} or simply \emph{tracing}. The
solutions are paths that manifestly conserve energy and angular momentum. The
integration may be formulated in a number of ways, namely as a second-order set
of ordinary differential equations (ODEs), analytically integrated once to give
a set of numerical first-order (ODEs), or twice to give position in terms of
elliptical integrals. The second-order ODE approach is conceptually the most
simple, unambiguously calculates the position and velocity of the geodesic at
each step of the integration, and is easiest to generalise to other spacetime
geometries. The second-order formulation is used in this paper.

Using coordinates $x^\mu$, the geodesic equation with external acceleration
$a^\mu$ ($=0$) is written
\begin{equation}
\label{eq:geodesic_equation}
    \frac{\d^2 x^\mu}{\d \lambda^2}
    + \utensor{\Gamma}{\mu}{\nu\sigma}
    \vel{\nu}
    \vel{\sigma}
    = a^\mu,
\end{equation}
where $\lambda$ is an affine parameter and $v^\mu := \d x^\mu / \d \lambda$ is
the four-velocity. The effects of spacetime curvature are encoded in the
Christoffel connection,
\begin{equation}
\label{eq:christoffel}
    \utensor{\Gamma}{\mu}{\nu\sigma}
    := \frac{1}{2} g^{\mu\rho}
    \left(
        \partial_{\nu}g_{\rho \sigma}
        + \partial_{\sigma}g_{\rho \nu}
        - \partial_{\rho}g_{\sigma \nu}
    \right),
\end{equation}
determined solely by the metric.

In this formulation, the geodesic equation is an initial value
problem that can be
solved with a choice of initial $x^\mu$ and $\vel{\mu}$. We are free to to
choose an initial 3-position $x^i$ (with $x^t = 0$ by convention), and constrain
the velocity using the geometric invariance
\begin{equation}
\label{eq:velocity_constraint}
    g_{\sigma\nu} \vel{\sigma} \vel{\nu} = -\mu^2,
\end{equation}
where $\mu$ is the invariant mass. This constraint gives rise to three solution
classes depending on the sign of $-\mu^2$; namely $-\mu^2 = 0$ corresponding to
\emph{null geodesics}, $-\mu^2 < 0$ to \emph{time-like geodesics}, and $-\mu^2 >
0$ to \emph{space-like geodesics}. Null geodesics are the trajectories of
photons through the spacetime, time-like geodesics are the trajectories of
massive particles, and space-like geodesics are the trajectories of exotic
`faster-than-light' particles.

As with position, it is sufficient to specify the three-vector $\vel{i}$,
and use \eqref{eq:velocity_constraint} to determine $\vel{t}$ by rearranging
\begin{equation}
\vel{t}  = \frac{-g_{t\phi} \vel{\phi} \pm
    \sqrt{-g_{ij} \vel{i} \vel{j} + \mu^2}
}{g_{tt}}.
\end{equation}
Sensible initial choices of $v^i$ will be discussed in the next section.

The positive and negative roots corresponds to the direction of time, wherein
lies the \textit{ray-tracing trick}: a time-reversal symmetry in the metric $t
\rightarrow -t$, $\phi \rightarrow -\phi$ allows geodesics to be calculated in
the forward direction as if they had travelled \textit{backwards}. Photons can
therefore be traced from an observer towards the black hole, but physical
calculations performed as if they had been emitted from near the black hole and
travelled back towards the observer. This seemingly trivial property cannot be
understated in an implementation of GRRT: it means only the `observed' geodesics
need to be traced, it will alter the directions of rotations when
applied, and inverts the energy ratio associated with points connected by
a particular geodesic.

Computing the geodesic equation requires some method of determining
$\utensor{\Gamma}{\mu}{\nu\sigma}$. This is usually analytically derived and
implemented laboriously by-hand. Sometimes computer algebra systems are used to
expedite the process, but this can lead to unnecessary computations in the final
expressions, or at the very least a high degree of code complexity that scales
poorly as the number of terms in the metric increases. Our alternative is to use
a numeric differentiation scheme, such as automatic differentiation (AD), to
calculate the Jacobian of the metric and therefore the Christoffel symbols
\emph{on-the-fly}\footnote{We were made
    aware by colleagues of the independently developed code Mahakala of
\citet{sharma_mahakala_2023} that uses a similar AD approach.}. If the
class of spacetime exhibits additional symmetries, these can be exploited to
reduce computation further: for example, metrics of the form
\eqref{eq:stationary_axisymmetric_metric} may exploit $\partial_t g_{\mu\nu} =
\partial_{\phi} g_{\mu\nu} = 0$ (the principal Killing vectors) and avoid
calculating two columns of the Jacobian entirely.

\subsection{Observers and emitters}
\label{sec:observers-and-emitters}

A different method is used to calculate the initial velocity vectors $v^\mu$
depending on whether an \emph{observer} or an \emph{emitter} is being
considered. This is for a convenience of parameterisation: observers are
represented by image planes, whereas emitters by their full $4 \pi$ solid angle
sky.

For an observer, we are interested in the sparse set of emitted photons which
reach an image plane representing the observer's field of view. We use a
`pinhole camera' setup, where every geodesic originates from the same point at
the observer, but with a velocity vector that has a slight angular offset in
order to intersect a particular point on the projected image plane. This is in
contrast to a setup where each `pixel' originates from an offset position, but
with parallel velocity vectors, mimicking a conventional image plane.

\begin{figure}
    \centering
    \includegraphics[width=0.99\columnwidth]{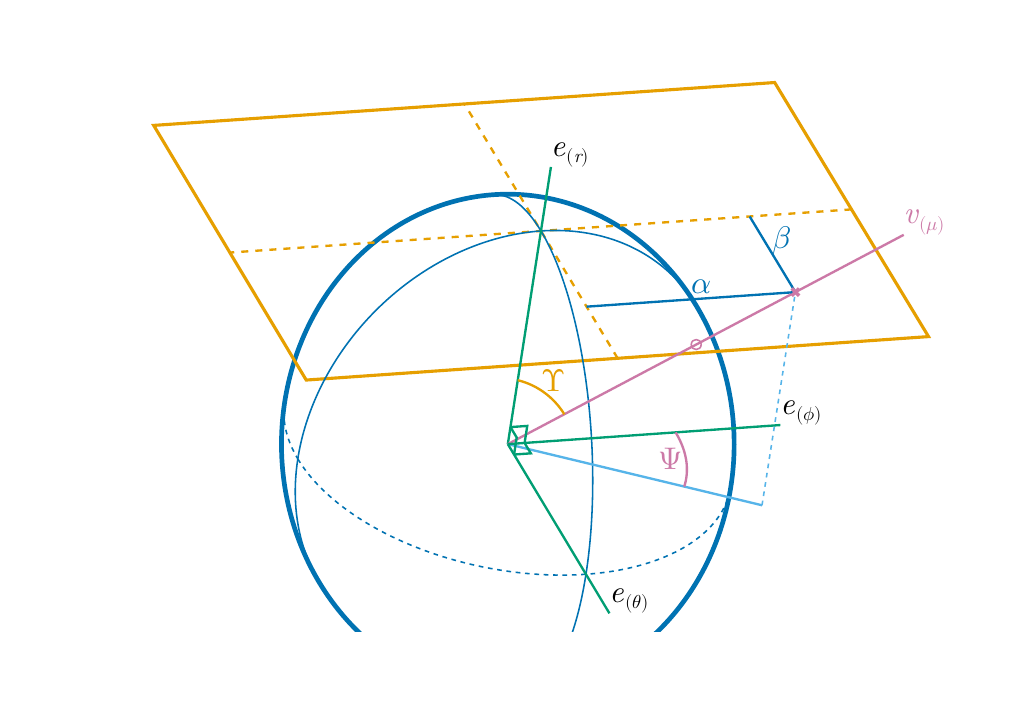}
    \caption{
    Geometry of the local sky for an observer or emitter. The image plane
    (yellow) is perpendicular to the $e_{(r)}$ axis, which, for an observer,
    points towards the the central singularity. The momenta $v_{(\phi)}$ and
    $v_{(\theta)}$ are used to calculate the impact parameters on the image
    plane, $\alpha$ and $\beta$ respectively. For emitters, the angles
    $(\Upsilon, \Psi)$ are used to parameterize vectors that point on the local
    sky, which may subsequently be decomposed onto the basis $e_{(i)}$ to find
    $v_{(\mu)}$.
    }
    \label{fig:observer-coordinates}
\end{figure}

The parameters representing the local sky are shown in
Fig.~\ref{fig:observer-coordinates}, with the projection of the image plane
drawn in yellow. Following \citet{cunningham_optical_1973}, one may then define
a set of \emph{impact parameters} from components of the local momenta $v_{(i)}$
\begin{align}
    \alpha &:=  x^r \frac{v_{(\phi)}}{v_{(r)}}, \\
    \beta &:= x^r \frac{v_{(\theta)}}{v_{(r)}},
\end{align}
where we have used indices in parentheses to denote the vector components in the
local frame. The choice of subscript ($r, \theta, \phi$) is to anticipate
identifying coordinate directions in the local and global bases, and to suggest
some intuition as to the meaning of directions in the local frame.

Exploiting the invariance of four-momenta, similar to
\eqref{eq:velocity_constraint}, we obtain a curve of solutions along $x^r$ for
the local momenta of geodesics intersecting the image plane at specific $\alpha$
and $\beta$
\begin{align}
    \frac{v_{(r)}}{v_{(t)}} &= -\left( \sqrt{1 +
    \left(\frac{\alpha}{x^r}\right)^2 + \left(\frac{\beta}{x^r}\right)^2}
\right)^{-1} = -\mathscr{R}, \\
    \frac{v_{(\theta)}}{v_{(t)}} &= \mathscr{R} \frac{\beta}{x^r}, \\
    \frac{v_{(\phi)}}{v_{(t)}} &= \mathscr{R} \frac{\alpha}{x^r}.
\end{align}
Note the choice of sign in the root of $v_{(r)} / v_{(t)}$, chosen so the
momentum points inwards towards the black hole.

The case for emitters is subtly different, where we now consider polar and
azimuthal angles in the local sky, denoted $\Upsilon$ and $\Psi$ respectively
(see also Fig.~\ref{fig:observer-coordinates}). The momenta components are
obtained by considering the tangent vector pointing along some initial direction
-- that is, by projecting the initial velocity vector onto the local momentum
frame. This projection is compactly expressed as a decomposition onto a
Cartesian coordinate system,
\begin{equation}
    \label{eq:local-angle-to-velocity}
    \frac{v_{(i)}}{v_{(t)}} = \frac{1}{v_{(t)}}
    \left[\pderiv{(x, y, z)}{(r, \theta, \phi)}\right]
    \left(
    \begin{matrix}
        \sin \Upsilon \cos \Psi \\
        \sin \Upsilon \sin \Psi \\
        \cos \Upsilon \\
    \end{matrix}
    \right),
\end{equation}
where the partial derivative matrix in square brackets denotes the Jacobian of
the Cartesian to spherical coordinate transformation. The constant of motion
component $v_{(t)}$ is defined as the negative of the energy in the local frame,
and $-v_{(t)}=E=1$ without loss of generality.

Directions and momenta are easiest to express in a local frame and then
transformed back to the global coordinates. The question arises which local
frame to use and the natural answer is the locally non-rotating frame
\citep[LNRF;][]{bardeen_rotating_1972}. This frame follows strictly circular
world lines, $x^r = \text{const.}$ and $x^\phi = \omega t + \text{const.}$,
with angular velocity $\omega = -g_{t\phi} / g_{\phi\phi}$. The transformation
from the LNRF is
\begin{equation}
    \label{eq:local-to-global-velocity}
    v_\mu = \e^{(\nu)}_{\phantom{(\nu)}\mu}\  v_{(\nu)}
\end{equation}
where the local tetrad (basis vectors) $\e^{(\nu)}_{\phantom{(\nu)}\mu}$ are
found using the theorem of Gram--Schmidt
\citep[][Appendix~\ref{appendix:gram-schmidt}]{schmidt_uber_1989}, similar to
\citet{wilkins_2020_isco}. The formalism
may be extended for a local frame in motion, where the frame velocity modifies
the mapping by a local Lorentz transformation,
$\Lambda^{(\kappa)}_{\phantom{(\kappa)}(\nu)}$, as
\begin{equation}
    v_\mu = \e^{(\nu)}_{\phantom{(\nu)}\mu}\  \Lambda^{(\kappa)}_{\phantom{(a)}(\nu)} v_{(\kappa)}.
\end{equation}
This Lorentz transformation may be absorbed into the tetrad basis, mandating the
velocity of the frame in the global coordinates to be defined as
$\utensor{\e}{(t)}{\mu}$ and orthogonalizing using the theorem of Gram--Schmidt.

These calculations may also be derived from the constants of motion of a
particular geodesic, namely energy $E$, angular momentum $L$, and the Carter
constant $Q$ \citep{carter_global_1968}. A derivation of the LNRF and associated
coordinate transformations is in \citet{cunningham_optical_1973}, where the
authors derive the impact parameters under the assumption that the observer is
in asymptotically flat space. Our calculations do not make this assumption, and
can therefore use impact parameters anywhere in the spacetime, approximating an
asymptotically flat space by positioning our observer at a large radial
distance, say $r_\text{obs} > 10^4 \rg$. This approximation incurs an error of
order $\sim1/r_\text{obs}$ when compared to results using asymptotic impact
parameters due to the energy and angular momentum differing by some small
relative to the asymptotic result.

\subsection{Special orbits and horizons}
\label{sec:special-orbits}

Of particular interest for studying accretion processes are the Keplerian
circular orbits. These are the circular orbits confined to the equatorial plane $(x^\theta = \pi/2)$
in the axis-symmetric spacetimes, and are used in a variety of accretion disc models
\citep{shakura_black_1973, sadowski_relativistic_2011}. They are stationary
points of the Hamiltonian of a geodesic, and may be solved from the geodesic
equation under the constraint $v^r = v^\theta = 0$. These are straightforward to
study analytically and have a general solution for metrics of the form
\eqref{eq:stationary_axisymmetric_metric} \citep[see e.g.][and
an extension towards $a^\mu \neq 0$ in Appendix
\ref{appendix:circular-orbits}]{johannsen_regular_2013}.

Circular orbits are classified as either stable or unstable depending on the
energetics of the orbit \citep{wilkins_bound_1972,bardeen_rotating_1972}. There
exists an innermost stable circular orbit radius (ISCO, denoted $\risco$), below
which circular orbits are energetically hyperbolic -- that is, small
perturbations will send test particles escaping to infinity or spiralling into
the central singularity. The stability of an orbit depends on the sign of $\d E
/ \d x^r$, with $>0$ corresponding to stable configurations. The ISCO is the
critical point
\begin{equation}
    \label{eq:isco-definition}
    0 = \left. \frac{\d E}{\d x^r} \right\rvert_{x^r := \risco}.
\end{equation}
Stable circular orbits are only possible for radii $x^r \geq \risco$. Within the
ISCO is the so-called \textit{plunging region} where $v^r \neq 0$. Emission from
within the ISCO is generally disregarded in reflection and reverberation models,
though for specific disc models and/or inner boundary torques emission from this
region may be important \citep[see e.g.][]{reynolds_isco_1997,young_isco_1998,
mummery_continuum_2024}. \Gradus makes no concrete assumptions about the
plunging region, though will omit contributions from within the ISCO in this
paper.

The ISCO radius may be solved numerically when no analytic solution is known.
For stationary, axis-symmetric metrics, the energy of a given orbit is
\eqref{eq:energy-of-orbit}, with the derivative with respect to $x^r$, which may
be calculated using a numerical technique. We use the NonlinearSolve.jl package
to solve for the root \eqref{eq:isco-definition} to high accuracy
\citep{Pal_NonlinearSolve_jl_2023}, using derivatives calculated with AD.

The event horizon radius is the radius of the coordinate singularity of the
spacetime. In lieu of an analytic formula, the horizon radius may be solved
using a similar numerical method as for the ISCO. For an axis-symmetry
spacetime, the event horizon is the set of $(r, \theta)$ coordinates that
satisfy
\begin{equation}
    \label{eq:event_horizon}
    0 = \left. \frac{1}{g_{rr}} \right\rvert_{x^r =: r_s},
\end{equation}
which may be solved assuming some convexity for $x^r$ given some $x^\theta$ and
$x^\phi$.

\subsection{Charts and horizons}

A \emph{chart} defines the boundaries of an integration domain that is used by
the integrator to terminate the computation when the fate of a given geodesic is
inescapable. In practical terms, it is a set of \emph{callback functions} in the
integrator that are evaluated at every time step to assess if the integration
should be terminated when a given condition has been met. They are used to
classify the outcome of an integration, e.g. whether a geodesic has fallen within the
horizon, escaped to infinity, intersected with the disc, and so on. In the
majority of cases, the inner and outer boundaries of the integration chart are
the event horizon $r_s$ and some \emph{effective infinity} $r_\infty$
respectively. This outer boundary denotes the region where a geodesic is assumed
to escape the potential of the singularity, or otherwise some arbitrary
outer cut-off radius for the integration depending on the problem being
simulated.

The geometry of an accretion disc or object is also expressed as a boundary of
the chart. In terminating the integration when the geodesic intersects such a
boundary the geodesic is said to have intersected the surface of the object,
whereupon it is labelled accordingly. These labels, termed \emph{status codes},
are used during analysis to filter or group geodesic solutions together.

Close to the inner radius the adaptive time step of certain ODE integration
algorithms will tends to zero due to near-singular derivatives, causing the
integration to slow almost to a standstill. We avoid this by scaling the inner
horizon with the choice of a constant $\mathcal{K} > 0$, such that $\tilde{r}_s
= (1 + \mathcal{K}) r_s$ is used to delineate the event horizon.  The constant
$\mathcal{K}$ may be adjusted depending on need. We set $\mathcal{K} = 10^{-2}$
(i.e. within 1\% of the true value) by default to balance performance and
accuracy.

\subsection{Observables}
\label{sec:computing-observables}

In \Gradus, an \textit{observable} is any physical quantity calculated from a
simulation. Computing an observable requires some or all pairs of $(x^\mu,
v^\mu)$ along a geodesic, though frequently only the start and end points are
necessary. An observable that requires multiple points along the geodesic can be
calculated either coincidentally with the geodesic equation, or subsequently
re-traced along the ray. Such quantities include parallel transport,
polarisation, or radiative transfer quantities. Calculating the quantities
simultaneously has the benefit that the error tolerance in the integrator is
sensitive to changes in the observable. The benefit of re-tracing is that for a
given set of metric parameters and observer positions, the trajectory of the
geodesic is unchanged, so the observable may be more efficiently recalculated.

A frequently used observables is the \emph{redshift}\footnote{Note that the term
`redshift' is in this paper refers to the redshift phenomena. It is used to
denote both the red-shift $g < 1$ and blue-shift $g > 1$.}, $g$. The redshift
associated with a geodesic connecting two points includes contributions from the
Doppler effect, special relativistic beaming, and gravitational redshift, and is
compactly written as the ratio of energies between the start and end point of a
geodesic,
\begin{equation}
\label{eq:redshift}
g := \frac{E_\text{end}}{E_\text{start}} = \frac{\left. v_\mu u^\mu
\right\rvert_\text{end}}{\left. v_\mu u^\mu \right\rvert_{\text{start}}},
\end{equation}
where $v_\mu$ are the geodesic (photon) momenta, and $u^\mu$ the velocity of the
emitting (start) and observing (end) media respectively.

\subsection{Disc illumination and emissivity profiles}
\label{sec:emissivity-profiles}

The presence of an illuminating source near the accretion disc, will give rise
to an \emph{illumination profile} $F_i$ over the accretion disc
\citep{svensson_corona_1994}. It is a function of the disc coordinates that maps
the flux deposited by the source in a patch of the disc. This is equivalently an
\emph{emissivity profile}, denoted $\varepsilon$, assuming the incident
radiation is back-scattered, and that the line strength in the reflected
spectrum is proportional to the illuminating flux, $\varepsilon = F_i$
\citep{wilkins_understanding_2012}. The incident radiation is photo-ionizes the
accretion disc and can therefore also be related to the ionisation parameter
$\xi$ of the accretion disc \citep{laor_line_1991,ross_effects_1993}, itself
given by
\begin{equation}
    \xi = \frac{4 \pi F_i (E)}{n_\text{H}}
\end{equation}
where $F_i (E)$ is the incident flux in a particular energy band, usually in the
range $0.01 - 100 \text{ keV}$, and $n_\text{H}$ is the co-moving hydrogen
number density \citep{ross_effects_1993}. This formulation assumes constant
density throughout the disc.

The illumination (and emissivity) profile is expressed as the ratio of source
solid angle area to disc area, or in the limiting case as the areas become
small,
\begin{equation}
    \varepsilon (r, \phi) = I \left\lvert \frac{\d \Omega}{\d A} \right\rvert,
\end{equation} \label{eq:emissivity}
up to a constant of normalisation.
Here $r$ and $\phi$ refer to coordinates on the surface of the accretion disc,
and $I$ is the specific intensity of the intrinsic coronal spectrum.

The geometry of the illuminating source -- the \emph{corona} -- is unknown. It
is often modelled, for simplicity, as a point source on the spin axis of the
black hole that emits isotropically. Such a configuration is known as the
`lamppost' corona \citep[e.g.][]{fukumura_accretion_2007}. For on-axis sources,
such as the lamppost, \eqref{eq:emissivity} is conveniently a function of $r$
only, and the calculations involved in determining the Jacobian term are greatly
simplified, as will be treated in a moment. Off-axis sources are more complex,
owing to both the two-dimensional nature of the emissivity profile and in how
the Jacobian term may be calculated. Other authors have explored off-axis
\emph{extended} geometries using Monte--Carlo codes
\citep{wilkins_understanding_2012, wilkins_towards_2016, gonzalez_probing_2017}.
Such an approach is computationally expensive, and can require some form of
dense tabulation. \Gradus has a novel formulation for computing fast
\emph{time-dependent emissivity profiles} for general off-axis extended coronae
that do not rely on Monte--Carlo methods \citep{baker_2025}.

For the lamppost corona the emissivity profile may be parameterised using only
the cylindrical radius on the disc, $\rho = x^r \sin(x^\theta)$. The $\lvert \d
\Omega / \d A \rvert$ term in in an annulus $\rho + \d \rho$ is approximated by
the number density of photons. The emissivity function is
\begin{equation}
    \varepsilon (\rho, \d \rho) = \frac{\mathcal{N}(\rho, \d \rho)}{\gamma
    \tilde{A}(\rho, \d \rho)} I(g),
\end{equation}
where $\mathcal{N}$ is the geodesic count in an annulus $\rho + \d \rho$, $I$ is
the intensity of the illuminating flux as a function of redshift $g$,
$\tilde{A}$ is the curvature corrected (proper) area of the annulus, and
$\gamma$ is the Lorentz factor that accounts for area contraction of the annulus
due to the velocity of the disc.

The relativistic corrections are as follows: for an infinitesimal area $\d A = \d
\rho\d\phi$, the \textit{proper area} is calculated directly from the metric,
such that
\begin{equation}
    \d\tilde{A} = \sqrt{g_{rr} g_{\phi\phi}}\, \d \rho\, \d \phi,
\end{equation}
is the area as measured by a stationary observer in the disc
\citep[e.g.][]{wilkins_understanding_2012}. The relativistic
Lorentz factor is calculated in the local frame
\begin{equation}
    \gamma = \frac{1}{\sqrt{1 - \left(v^{(i)}\right)^2}},
\end{equation}
where $v^{(i)}$ are the spatial components of the angular velocity in the LNRF.
These velocity components are determined for a given basis
\begin{equation}
    v^{(i)} = \frac{\utensor{\e}{(i)}{\mu}\, v^\mu}{\utensor{\e}{(t)}{\sigma}\,
    v^\sigma}.
\end{equation}
For the special case of circular orbits in the equatorial plane, only
$v^{(\phi)}$ in this equation is non-zero.

The emission spectrum for the illuminating corona is usually assumed to be a
powerlaw $I(g) = g^{-\Gamma}$ with photon index $\Gamma$. For many applications
$\Gamma$ is assumed to be $1 - 3$
\citep{mushotzky_agn_pl_1982,remillard_binaries_2006}. For emission to be
locally isotropic means the geodesics from the source are traced by sampling the
sky angles $\Upsilon$, $\Psi$ evenly on a sphere (see Figure
\ref{fig:observer-coordinates}). The angles are transformed via Equations
\eqref{eq:local-angle-to-velocity} and \eqref{eq:local-to-global-velocity} to
find the initial velocities of the geodesics, which are then integrated to solve
for their trajectories. When the emission is locally non-isotropic, the sampled
distributions of $\Upsilon$, $\Psi$ may be appropriately weighted to give the
desired pattern.

\begin{figure}
    \centering
    \includegraphics[width=0.95\columnwidth]{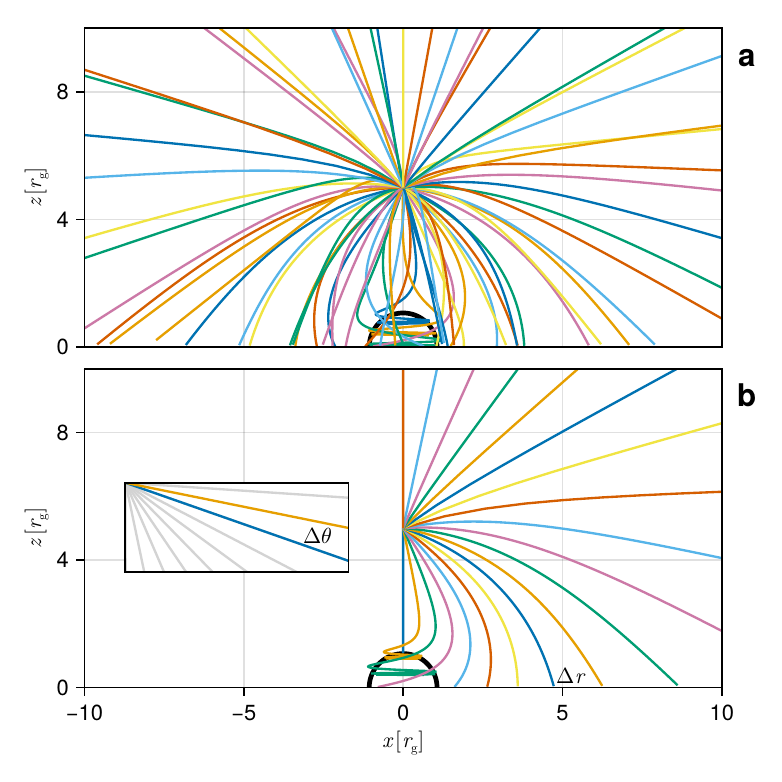}
    \caption{An illustration of the methods described in the text for
        calculating the emissivity of a lamppost corona: the colourful lines are
        the null-geodesics of a lamppost model illuminating the accretion disc
        around in a maximally spinning Kerr spacetime ($a = 0.998$). Panel a) is
        the Monte--Carlo approach, where the initial velocity vector of the
        photon is sampled isotropically on the local sky of the emitter. The
        number density on the disc in a given annulus is then used as a proxy
        for the flux density. Panel b) shows how the symmetry of the lamppost can
        be exploited, by considering only a slice of the emission around the
        spin axis. The initial velocity vectors now differ by a constant $\Delta
    \theta$, which allows the spacing on the disc $\Delta r$ to be used as a
proxy for flux density.}
    \label{fig:coronal-tracing}
\end{figure}

The lamppost emissivity profile may alternatively be determined by exploiting
symmetries as in \citet{dauser_irradiation_2013}.  Here, only a given slice with
constant $\Psi$ is considered, where $\Upsilon$ is now evenly spaced. Pairs of
geodesics separated by some small $\Delta \Upsilon$ have endpoints on the disc
separated by $\Delta \rho$, such that $\sin (\Upsilon) \Delta \Upsilon / \Delta
\rho \propto \mathcal{N}$ can be used as a proxy for the number
density\footnote{The $\sin \Upsilon$ factor comes from the distribution being
    isotropic. One may equivalently use $1/\Delta \rho \propto \mathcal{N}$ and
    sample $\Upsilon \sim \cos (1 - 2 \mathcal{U})$, where $\mathcal{U}$ is a
    uniform distribution $\mathcal{U}(0,1)$. This result may be shown using the
inverse-CDF or Smirnov transform method.}. The emissivity is then written
$\varepsilon$ is weighted similarly,
\begin{equation}
    \varepsilon(r, \Delta \rho) = \frac{\sin \Upsilon}{\gamma \tilde{A}(\rho)}
    \frac{\Delta \Upsilon}{\Delta \rho} I(g).
\end{equation}
A conceptual illustration of the two methods is shown in Figure
\ref{fig:coronal-tracing}. The latter method converges much faster than the
former Monte--Carlo sampling approach, however is only valid for on-axis coronal
models.

\subsection{Transfer functions}
\label{sec:transfer-functions}

\begin{figure*}
    \centering
    \includegraphics[width=0.95\linewidth]{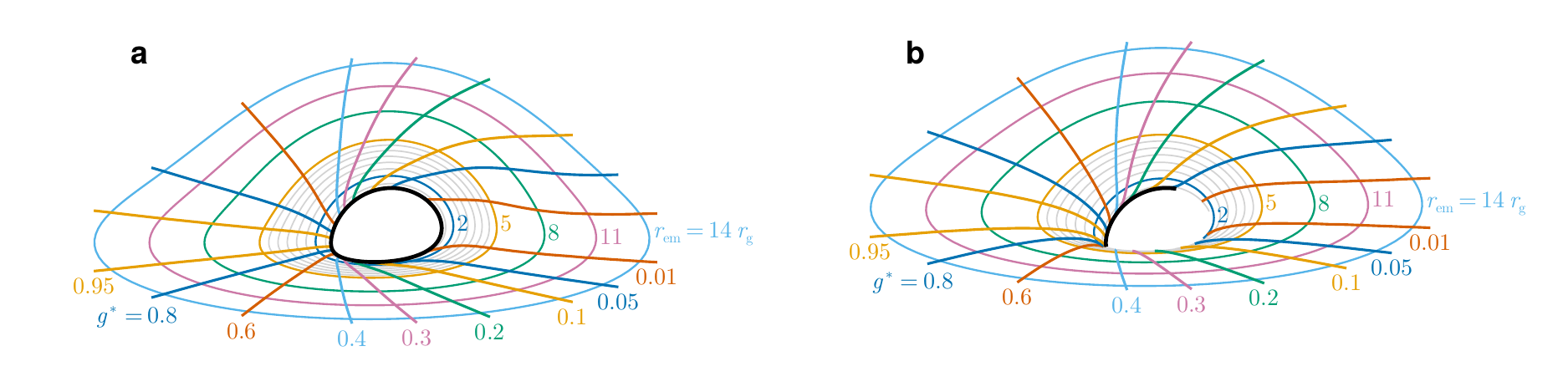}
    \caption{Concentric rings of radius $\rhoem$ and contours of constant
    dimensionless redshift $g^\ast$ on disc in the equatorial plane, projected
on the image plane of a distant observer at $\theta_\text{obs} = 75^\circ$. The
central singularity is described by the Kerr metric with $a = 0.998$. The
innermost thick black line is the projection of the ISCO. Note the contours of
$g^\ast$ are double valued for any given $\rhoem$. Panel a) geometrically
thin disc. Panel b) \citet{shakura_black_1973} Disc (SSD) with $\dot{M} / \dot{M}_\text{Edd} = 0.3$, with
obscuration of some of the inner radii. The edge of the ISCO is here only
partially visible.}
    \label{fig:transfer-parameterisation}
\end{figure*}

To motivate the use of \emph{transfer functions}, consider the means by which an
observed spectrum of flux reflected by the accretion disc is calculated. The infinitesimal
flux element, $\d F(E)$, is related to the specific intensity in the solid angle
element $\d \Omega$ by
\begin{equation}
    \label{eq:infinitesimal-flux}
    \d F(E) = I(E)\, \d \Omega,
\end{equation}
for observed energy $E$ and specific intensity $I (E)$. The relation between the
observed and emitted intensities is derived using the \emph{reciprocity theorem}
\citep[equivalently \emph{Liouville's theorem},][]{lindquist_louville_1966} with
\begin{equation}
\label{eq:liouville-theorem}
    I_\text{obs}\left( E_\text{obs}\right) = g^3 I_\text{em}\left(E_\text{em}\right).
\end{equation}
For the intuition behind this relation, we refer the reader to
\citet{ingram_public_2019}.

To obtain the flux as measured by the observer, an integration over the
observer's field of view, $\d \Omega$, is performed. Up to a constant, this is
equivalent to integrating over the image plane $\d \alpha \d \beta$,
\begin{equation}
\label{eq:integrate-impact-params}
F(E_\text{obs}) \propto \int I_\text{obs}(E_\text{obs}) \d \alpha \d \beta.
\end{equation}
In practical terms, this is simply binning the flux element associated with each
geodesic in each pixel on the image plane. Formulating the flux calculation in
this way is computationally simple but expensive, as the flux is computed on a
pixel-by-pixel (geodesic-by-geodesic) basis. The emitted intensity $I_\text{em}$
depends on the disc emissivity, requiring both coordinate and redshift values
for each geodesic to be stored in memory. These quantities are dependent on the
metric, disc, and observer parameters, and this dependence limits the reuse of
calculations between simulations -- an undesirable property for a portable
spectral model.

To avoid such problems, the relativistic effects in the flux calculation are
encoded in so-called \emph{transfer functions}
\citep[e.g.][]{brenneman_constraining_2006}. The most ubiquitous formulation of
these transfer functions was first introduced in \citet{cunningham_effects_1975}.
There, they are defined
\begin{equation}
    \label{eq:cunn-transfer-function}
    f:=\frac{g}{\pi \rhoem} \sqrt{g^\ast(1 - g^\ast)} \jacobian{(\alpha, \beta)}{(\rhoem, g^\ast)},
\end{equation}
where $\rhoem$ is the emission radius on the disc, and
\begin{equation}
    g^\ast := \frac{g - g_\text{min}}{g_\text{max} - g_\text{min}} \in [0, 1],
\end{equation}
is a rescaled dimensionless redshift parameter which acts as a proxy of the
$\phi$ coordinate on the disc. The extremal values of $g$ are
calculated for a given emission radius on the disc, $\rhoem$, and so
$g_\text{min}$ and $g_\text{max}$ are implicitly functions of $\rhoem$
themselves.

The parameterization of $g^\ast$ is double-valued everywhere except at $g^\ast =
0$ and $1$, with the two values of $g^\ast$ being attributed to the points along
$\rhoem$ that are closest and furthest from the observer (see
Figure~\ref{fig:transfer-parameterisation}). This leads to an interpretation of the
Cunningham transfer functions as recasting the projection of the accretion disc
on the image plane from $(\alpha, \beta)$ to $(\rhoem, g^\ast)$, see
Figure~\ref{fig:transfer-parameterisation}. The additional elliptical envelope in
$g^\ast$ suppresses the singular values of the Jacobian as $g^\ast$ becomes $0$
or $1$, resulting in numerically better behaved functions at extremal $g^\ast$
that can be integrated over $g$.

As a note, the name \emph{transfer functions} is a more general term and used
elsewhere, for example in reverberation lags discussed later. To avoid
ambiguity, we henceforth refer to functions of the kind defined in
Equation~\eqref{eq:cunn-transfer-function} as \emph{relativistic} or
\emph{Cunningham transfer functions}. Furthermore, the Cunningham transfer
functions are referred to as having `upper' and `lower' branches between
extremal $g^\ast$, as shown in Figure~\ref{fig:transfer-sampling-pattern},
stemming from the double-valued nature of $g^\ast$. When the emission from the
disc is not isotropic, the distinction between these branches becomes important,
as the geodesic that connects the disc element to the observer makes a different
cosine angle to the disc surface.

\subsubsection{Calculating Cunningham transfer functions}

Our method for calculating $f$ is a variation of the algorithms of other authors
\citep{speith_photon_1995,bambi_testing_2017,abdikamalov_public_2019}, and uses
AD to compute the Jacobian term. The procedure is as follows.

First, we find the curve of impact parameters that map to a ring of radius
$\rhoem$ on the accretion disc. This curve is the boundary of a star-convex
region on the image plane, and can therefore be written as
$\mathcal{R}(\vartheta)$, with $\alpha = \mathcal{R}(\vartheta) \cos(\vartheta)$
and $\beta = \mathcal{R}(\vartheta) \sin(\vartheta)$. For a given $\vartheta$,
the offset on the image plane $\mathcal{R}$ is found by root-finding the
difference between $\rhoem$ and the projected endpoint of the geodesic on
the disc $\rho = x^r (\mathcal{R}) \sin x^\theta(\mathcal{R})$, using a novel
root-finding method that takes advantage of the AD derivatives that can be
calculated along a geodesic. The method is principally a Newton--Raphson
implementation with an additional pivot point that is used to estimate a
bracketing interval, used if the Newton-Raphson method should fail. In such
cases, the bracket is used with the Golden--Section implementation of
\citep{Optim.jl-2018}.

Second, we must determine the extrema of $g$ over the $\rhoem$ ring. From the
previously calculated $\vartheta$, the best estimate of the extrema of $g$ is
used as a starting point for the same Golden--Section bracketing method, where
each step of the extremiser calls the root-solving implementation to find the
$\mathcal{R}(\vartheta)$ that intersects $\rhoem$. Since these extrema are
usually close to the line along $\beta = 0$, we shift the domain of the problem
to $\vartheta \in [ -\pi/2, 3\pi/4)$, which helps ensure the maxima and minima
are away from the edges of the domain. This helps to construct contiguous
brackets for the optimisers.

The importance of accurately calculating $g_\text{min}$ and $g_\text{max}$ is
difficult to overstate: small errors here will dramatically alter the shape of
the transfer functions close to $g^\ast \rightarrow 0$ and $g^\ast \rightarrow
1$, and have a high likelihood of sending $f$ to positive or negative infinity.
Even when the extrema are determined to high accuracy, in practice, it is useful
to employ a small truncation either side of the domain, as is discussed in the
next section in the context of integrating $f$.

Finally, the Jacobian is calculated by retracing all previously traced $(\alpha,
\beta)$ that map to $\rhoem$ with AD. Other authors will here use
\begin{equation}
    \left\lvert
    \pderiv{(\alpha, \beta)}{(\rhoem, g^\ast)}
    \right\rvert
    =
    \left\lvert
    \pderiv{\rhoem}{\alpha}\pderiv{g^\ast}{\beta}
    -
    \pderiv{\rhoem}{\beta}\pderiv{g^\ast}{\alpha}
    \right\rvert^{-1},
\end{equation}
and determine the derivatives with a finite difference stenciling approach, or
even use a fixed $\delta \alpha$ and $\delta \beta$. Unless the algorithm can
adapt extremely well, this approach may introduce singular values or risk large
numerical error at extremal $g$. AD avoids these problems, and has the
additional advantage that it only requires the evaluation of a single geodesic
to compute.

\begin{figure}
    \centering
    \includegraphics[width=0.95\columnwidth]{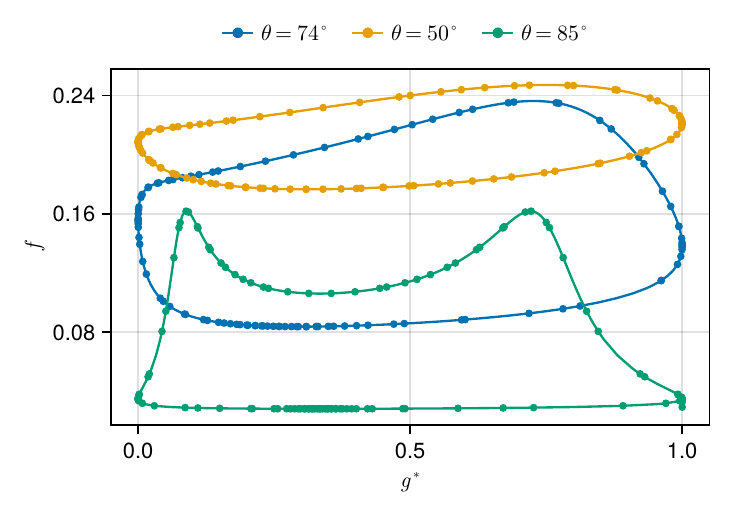}
    \caption{Transfer functions of the Kerr spacetime ($a = 0.998$) for various
        observer inclinations, showing the sample pattern of $g^\ast$ that our
        algorithm (described in the text) produces. The higher density of points
        around extremal $g^\ast$ is due to the optimiser
        that determines the extremal points. Note that for the high inclination
        case ($\theta = 85^\circ$) there is slight numerical noise very close to
        $g^\ast = 1$. This noise is in the region excluded by the integration
        scheme, and therefore contributes negligible error in calculations
        involving transfer functions. There is also a high density of points on
        the lower branch of the transfer function, which is an artefact of the
        projection of the disc onto the observer's plane.
    }
    \label{fig:transfer-sampling-pattern}
\end{figure}

The total number of geodesics traced for each Cunningham transfer function
depends on the number of steps needed to solve impact parameters for a given
$\rhoem$, and on the number of steps needed to extremize $g^\ast$ to within
some tolerance. In our code, we set an upper-limit on the number of steps so
that memory can be contiguously allocated, at the cost of possibly curtailing
some calculations before tolerance is reached. Our default configurations use an
upper limit of $N = 114$ points along $\rhoem$, $80$ of which are
approximately linearly sampled in $\vartheta$, and $34 = 2 \times 17$ used to
determine $g_\text{min}$ and $g_\text{max}$. These limits were chosen to balance
speed and accuracy of calculation. The resulting sampling pattern is sensitive
to extrema in $g^\ast$, as shown in Figure~\ref{fig:transfer-sampling-pattern}.

\subsubsection{Partially obscured Cunningham transfer functions}
\label{sec:partially-obscured-functions}

For thick accretion discs, it is possible that the disc obscures itself, and
that certain $\rhoem$ may therefore be only partially visible or entirely
obscured to an observer at inclination $\theta_\text{obs}$ (see
Figure~\ref{fig:transfer-parameterisation}b). When this is the case, the method
of calculating Cunningham transfer functions is modified through the use of
\emph{datum planes}.

\begin{figure}
    \centering
    \includegraphics[width=0.95\columnwidth]{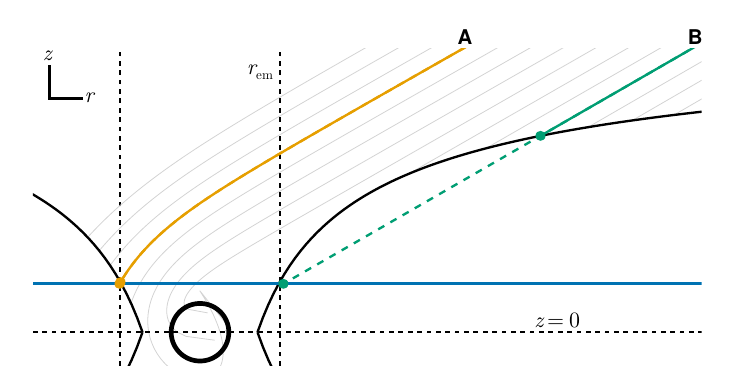}
    \caption{Cross-sectional slice of geodesics traced from a distant observer
        against a datum plane (horizontal blue line) through a thick disc
        (curved black line). The observer `sees' the solid geodesics
        intersecting the disc, whereas for the purposes of calculating the
        Cunningham transfer functions, we continue integrating along the dashed
        line until intersecting the datum plane. For geodesic A the observer
        can see the emission radius $\rhoem$, whereas for B the radius is
        obscured.
}
    \label{fig:datum-plane-tracing}
\end{figure}

We define a \emph{datum plane} to be an infinite plane at some scale height
$z_\text{em} = h(\rhoem)$, where $h$ is the height of the disc above the
equatorial plane. The datum plane is used as the manifold over which the
optimiser solves the projection of the ring with radius $\rhoem$. The datum
plane is in essence a cut-off for integration such that all geodesics terminate
at the same height, see Figure~\ref{fig:datum-plane-tracing}.

\begin{figure*}
    \centering
    \includegraphics[width=0.99\linewidth]{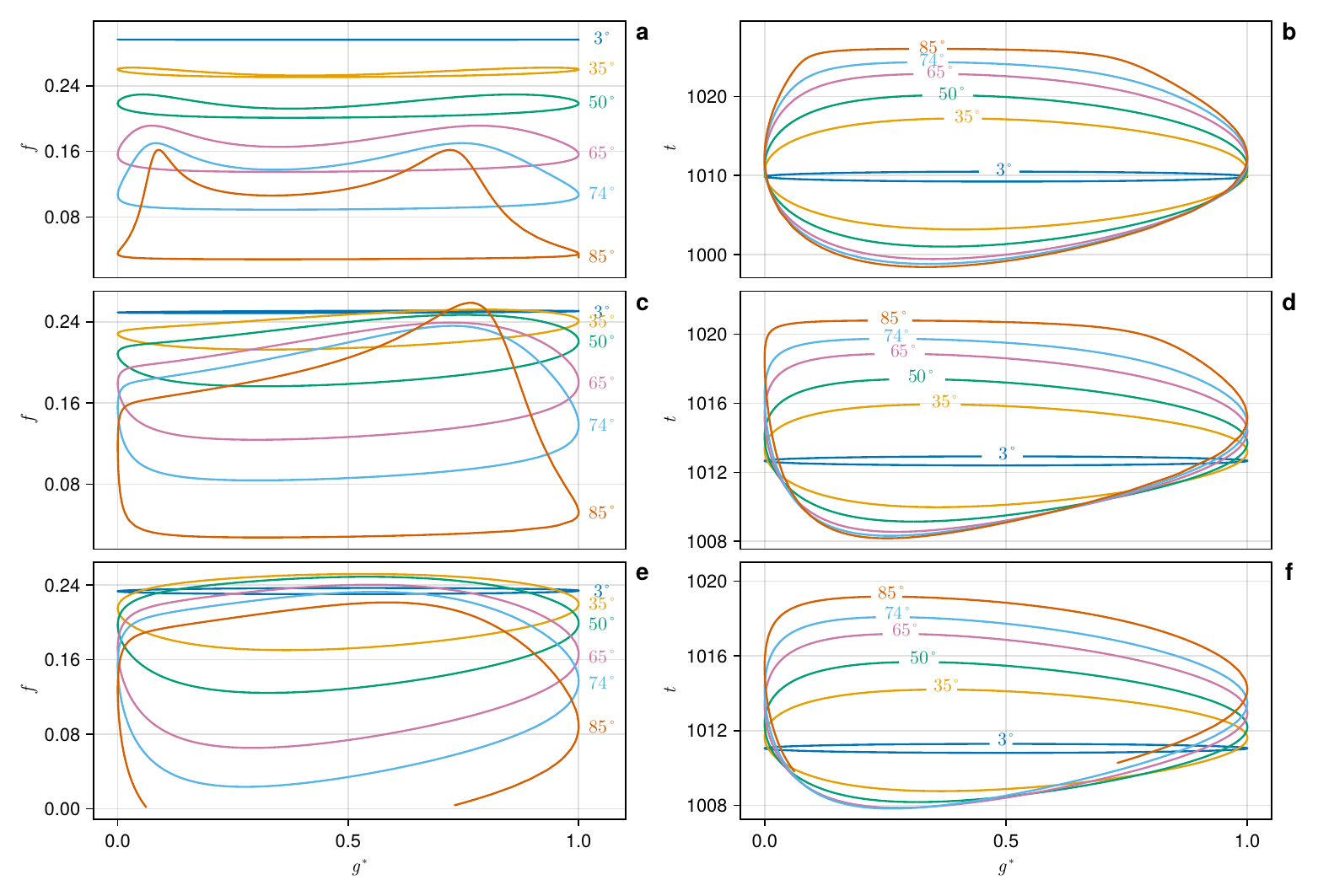}
    \caption{Transfer functions $f$ and timing $t$ for fixed $\rhoem$ for
        different observer inclinations $\theta_\text{obs}$ and $r_\text{obs} =
        10^3 \rg$. Panels a) and b) Schwarzschild spacetime with equatorial thin
        disc and $\rhoem = 11\, \rg$. Panels c) and d) Maximally spinning Kerr
        spacetime $a=0.998$ with equatorial thin disc and $\rhoem = 4 \, \rg$
        \citep[see also][their Figure 1]{bambi_testing_2017}. Panel e) and f)
        Maximally spinning Kerr spacetime $a=0.998$, with SSD $\dot{M} /
        \dot{M}_\text{Edd} = 0.3$ and $\rhoem = 4\, \rg$, showing obscuration at steep
        inclination.}
    \label{fig:transfer-functions}
\end{figure*}

As before, we determine the extremal redshift over the ring $\rhoem$, including
those geodesics from obscured regions of the disc. As the redshift depends only
on the start and end point of the geodesic, it is therefore simpler to combine
this with the radius-solving step and calculate the redshift over the datum
plane. This is not so for the Jacobian term, which depends on the cross section
traced by a bundle of geodesics for which the geometry of the disc is important.
The Jacobian is calulated by tracing against the thick disc and is combined with
a check to see if a patch of the disc is obscured. This obscuration check is
used to mask values of the transfer functions, as in
Figure~\ref{fig:transfer-functions}. This method is in effect analogous to the
`imaginary photons' of \citet{abdikamalov_testing_2020}, formalized for our
specific approach and the use of AD.

\subsection{Integrating transfer functions}
\label{sec:transfer-function-integration}

Using the Cunningham transfer functions we can rewrite the integration
\eqref{eq:integrate-impact-params} as
\begin{equation}
    F(E_\text{obs}) = \int \frac{\pi \rhoem f(\rhoem, g^\ast)}{g \sqrt{g^\ast (1 - g^\ast)}}
    I_\text{obs}(E_\text{obs}) \ \d \rhoem \d g^\ast,
\end{equation}
where we can use reciprocity theorem \eqref{eq:liouville-theorem} to express the
observed intensity in terms of the emitted intensity, and write $E_\text{obs} =
E_\text{line}g$ . The integrand is then only a function of $\rhoem$ and
$g^\ast$. A new interpretation of the Cunningham transfer functions is then
evident, namely as the Green's function of an annulus $\rhoem$ in response to
some delta function of $I_\text{em}$. This interpretation is used to guide
integration schemes for the transfer functions \citep{dauser_broad_2010}.

The Cunningham transfer functions are split into the aforementioned upper and
lower branches at extremal $g^\ast$, and each branch is integrated separately
and summed in the final result. We extend the transfer function integration to
include the geodesic coordinate time, tabulated as part of the transfer function
calculation. To include this timing information in the observed flux, we write
the integral
\begin{align}
    \label{eq:transfer-integration}
    F(E, t) &=
    \pi
    \int_0^\infty \d t^\prime \delta(t - t^\prime)
    \int_{r_\text{in}}^{r_\text{out}} \d \rhoem\,\rhoem \nonumber \\
    &\ \int_0^1 \d g^\ast\, \delta(E - gE_\text{line})\, g^3 I_\text{em}\left(\frac{E}{g}, t^\prime\right) \frac{f(\rhoem, g^\ast)}{\sqrt{g^\ast (1 - g^\ast)}},
\end{align}
denoting $g = g( \rhoem, g^\ast, t')$ implicitly. The Dirac delta functions
select the $E$ and $t$ bins of the observed flux respectively. We note that this
integrand is difficult to evaluate, and that the notation serves only to
compactly express the mathematical formalism of the integration. In practical
terms, we have a discrete set of Cunningham transfer functions that can be
smoothly interpolated. The method for numerically integrating the time-dependent
transfer functions pre-allocates an output matrix with time on one axis and
energy on the other. After choosing the radial limits of the integral
$r_\text{in}, r_\text{out}$, the total energy range over the matrix,
$E_\text{min}, E_\text{max}$ and time range $t_\text{min}, t_\text{max}$ are
determined. The method is then to evaluate the integral at a fixed (small)
energy and time resolution, determined by the limits of the transfer function,
and to sum each (small) contribution into the corresponding element of the
output matrix.  Any evaluation of \eqref{eq:transfer-integration} that is
outside of the energy and time limits is discarded, and therefore the axes of
the matrix must be chosen with some degree of care.

The integration over $\d \rhoem$ is performed using a simple trapezoidal scheme,
as it is sufficiently accurate and performant. The integration over $\d g^\ast$
requires more care, particularly around the extremal values, for which we use a
7$^\text{th}$ order Gauss--Kronrod quadrature integration scheme. The flux for a
particular energy $F(E)$ is evaluated for each branch over a small output bin in
$g$, whereas $t$ is determined by evaluating $t(g^\ast)$ at the limits of the
bin. Depending on the timing resolution required, is it imperative that the
integration is performed over a \emph{fine} $g$ grid. The method in full is:
\begin{enumerate}
    \item Interpolate the values of the transfer function, $f$, $t$, and
        $g_\text{min}$, and $g_\text{max}$, for the current radius $\rhoem$.
    \item Calculate the trapezoidal integration weighting for the radial
        coordinate $\omega_i = \Delta \rhoem \rhoem$. Any other quantities that only
        depend on $\rhoem$ may similarly be factored into this weighting. The
        emissivity function can be included here, should it depend only on
        emission radius, i.e. $I_\text{em}(\rhoem)$.
    \item Construct the fine $g$ grid from $E_\text{min}$, $E_\text{max}$, using
        $g = E / E_\text{line}$. The only requirement for this grid is that it
        is has at least twice the resolution of the output matrix to avoid
        artifacts of interpolating $f$. This can be increased for better
        resolution at computational cost.
    \item For each bin in the fine $g$ grid, integrate both branches of $f$ over
        this bin. Record $t_\text{min} = t(g_\text{min})$ and $t_\text{max} =
        t(g_\text{max})$. Caution must be taken, as depending on the
        interpolation scheme or implementation, if $f$ is a function of $g$
        instead of $g^\ast$, a $g / (g_\text{max} - g_\text{min})$
        change-of-variable factor must be included in the integrand.
    \item Add $\omega_i F_i(E, t)$ to the bin corresponding to the energy $E =
        gE_\text{line}$ and time $t$. If the range $g_\text{min}$ to
        $g_\text{max}$ straddles an output bin in either energy or time, the
        flux components in each bin must be weighted appropriately and summed
        into their corresponding bins.
\end{enumerate}

The integrand for each branch may still in practice diverge at $g^\ast
\rightarrow (0, 1)$. The above integration is therefore only performed with
support $g^\ast \in [h, 1 - h]$. Outside of this domain, the limits of the
integrand can be taken to approximate the edges of the bin, as in
\citet{dauser_broad_2010},
\begin{equation}
    F_\text{edge}(E,t) \propto 2\left( \sqrt{E_\text{max}} - \sqrt{E_\text{min}} \right).
\end{equation}
The constant of proportionality is determined by evaluating the integrand at $h$
or $1 - h$ respectively.

We use $h = 2 \times 10^{-8}$. The grid in $\rhoem$ over which the transfer
functions are interpolated should be irregularly spaced, e.g.  $\sim 1 / r$ or
as a geometric series, reflecting the fact that the majority of the variation in
$f$ occurs at small $r$. For $g^\ast$, the variation is approximately uniform
over the domain, and so we use a uniform sample of $N = 30$ points in $[0,1]$.
This also has the advantage that it makes the transfer functions simple to
export in tabular format.

Formulating the Cunningham transfer function integration with a time component
allows the same transfer function table to be used in computing both spectral
and timing properties. The integration can be performed sufficiently fast for
use with parameter inference. We also note that extensions to $I_\text{em}$ that
require, for example, the photon emission angle relative to the disc, such as
limb darkening or fluorescence \citep{matt_reflection_1993}, are straightforward
to include. They require, however, that each transfer function branch is
integrated separately, as previously noted.

\subsection{Covariant radiative transfer}

The intensity of a geodesic is an \emph{observable} that depends on every point
along the geodesic. The covariant formulation of the radiative transfer equation
calculates the emissions and extinction in a frame co-moving with the geodesic
\citep{fuerst_radiation_2004,younsi_general_2012}. The generalized form of the
differential equation with respect to the affine parameter $\lambda$ can be
written in terms of the Lorentz invariant intensity $\mathcal{I} = I_\nu /
\nu^3$ \citep{lindquist_louville_1966} as follows,
\begin{equation}
    \label{eq:covariant-radiative-transfer}
    \frac{\d \mathcal{I}}{\d \lambda} = \left. \frac{\d s}{\d \lambda} \right\rvert_\lambda \left( -\alpha_\nu \mathcal{I} + \frac{j_\nu}{\nu^3} \right),
\end{equation}
where $\mathcal{I}$ is the invariant intensity, $s$ is the proper length
traversed by the geodesic, and $\alpha_\nu$ and $j_\nu$ are the frequency $\nu$
dependent absorption and emissivity coefficients respectively, as measured in
the local frame. The frequency $\nu$ is related to the observed frequency via
the redshift,
\begin{equation}
    \nu = \frac{\nu_\text{obs}}{g}.
\end{equation}
In general, both coefficients $\alpha_\nu$ and $j_\nu$ are fields that depend on
the position $x^\mu$.

The $\d s / \d \lambda$ derivative term is calculated by projecting the geodesic
momentum $v_\mu$ onto the velocity $u^\mu$ of the medium, using the projection
tensor
\begin{equation}
    \mathrm{P}^{\mu\nu} := g^{\mu\nu} + u^\mu u^\nu.
\end{equation}
The path length derivative is
\begin{align}
    \left. \frac{\d s}{\d \lambda} \right\rvert_\lambda
    &= - \left. \left\lVert \mathrm{P}^{\mu\nu} v_\mu\right\rVert\, \right\rvert_\lambda,\\
    &= - \left. \sqrt{v_\mu v^\mu + \left(v_\mu u^\mu\right)^2 \left(2 + u^\mu u_\mu\right)} \, \right\rvert_\lambda,
\end{align}
such that for the particular case of null geodesics through a time-like medium
\begin{equation}
    \left. \frac{\d s}{\d \lambda} \right\rvert_\lambda = - \left. v_\mu u^\mu \right\rvert_\lambda.
\end{equation}
The intensity is calculated by selecting $\nu_\text{obs} = E$ at the observer,
and integrating \eqref{eq:covariant-radiative-transfer} along a given geodesic.

%%% DESCRIPTION OF THE CODE %%%%%%%%%%%%%%%%%%%%%%
\section{Description of the code}
\label{sec:description-of-code}

\Gradus is open-source and publicly available. It is implemented in the Julia
programming language \citep{Bezanson_Julia_A_fresh_2017}. We use the
DifferentialEquations.jl ODE solving library with ForwardDiff.jl for
forward-mode automatic differentiation
\citep{rackauckas_differential_2017,RevelsLubinPapamarkou2016}. The code is
available via the \software{Pkg} Julia package manger in a registry maintained
by the University of Bristol astrophysics
group\footnote{\url{https://github.com/astro-group-bristol/AstroRegistry/}}. The
software supports multi-threaded execution for CPUs, with limited GPU support.

\Gradus has a single expressive high-level programming interface for a variety
of GRRT problems, with sensible defaults and optional fine-grained control. The
software is extensively tested with a suite of unit and integration tests. The
tests are constructed both by comparing numerical algorithms to specific
analytic counterparts, and by comparing against snapshots of results in the
literature. \Gradus implements many algorithms for solving the same problem,
providing the choice of which method to use depending on the context. The
multitude of algorithms are also used as a self-consistency test, to ensure the
validity of our various implementations.

\Gradus aims to make exploring new spacetime models simple by only requiring the
non-zero metric components to be implemented. To this end, and for comparison, we
maintain a catalogue of metric implementations, including the Kerr spacetime,
Morris--Thorne wormhole \citep{morris_wormholes_1988}, Johannsen metric
\citep{johannsen_regular_2013}, the Einstein--Maxwell--Dilaton--Axion metric
\citep{garcia_class_1995}, the No-$\mathbb{Z}_2$ metric
\citep{chen_observational_2024}, and the Kerr--Newman
metric \citep[e.g.][]{hackmann_charged_2013}, complete with the ability to
specify the electromagnetic potential vector from which external accelerations
$a^\mu$ in Equation~\eqref{eq:geodesic_equation} are calculated.

The design of \Gradus prioritizes usability and extensibility, which comes at a
small performance cost: our aim is not to implement the fastest, semi-analytic
solutions to specific problems, but rather to have an optimal and interpretable
codebase for exploring problems related to general relativity.  This is not to
imply \Gradus is slow: our geodesic integration is sufficiently fast, and the
use of AD and optimisers means our algorithms solve problems fast enough to
merit no more than a personal computer. The abstractions in \Gradus have been
designed to allow users to implement and calculate observables of their models
quickly. To this end, we also include a number of visualization and plotting
recipes to provide some intuition for the problem space.

This design is possible with Julia's just-in-time compilation and multiple
dispatch, which also brings additional benefits: different number types may be used
through the whole library, such as arbitrary precision floating point
operations, symbolic types through Symbolics.jl \citep{symbolics_julia}, or the
propagation of AD gradient information through an entire simulation.
Consequently, our ray-tracer is entirely differentiable, which we make use of
through gradient-enhanced optimisers where mathematical optimisation is needed.

%%% TEST PROBLEMS %%%%%%%%%%%%%%%%%%%%%%%%%%%%%%%%
\section{Test problems}
\label{sec:test-problems}

In this section we validate our code using a number of standard test cases from
the literature. Unless otherwise stated, all test problems are integrated with
the Tsitouras 5-4 algorithm \citep{tsitouras_rungekutta_2011}. We also consider
Faegin's 10$^\text{th}$ order Runge--Kutta method, the explicit Runge--Kutta
4$^\text{th}$ order algorithm \citep{press_numerical_2007}, and Verner's `Most
Efficient' 6-5 Runge--Kutta method.  These are all implemented in the Julia
DifferentialEquations.jl package \citep{rackauckas_differential_2017}, and used
to highlight and small numerical differences that accompany the choice of ODE
integrator.

Many of the test problems are taken from \citet{gold_verification_2020} for ease
of comparison.

\subsection{Integration accuracy and stability}

\begin{figure}
    \centering
    \includegraphics[width=0.95\columnwidth]{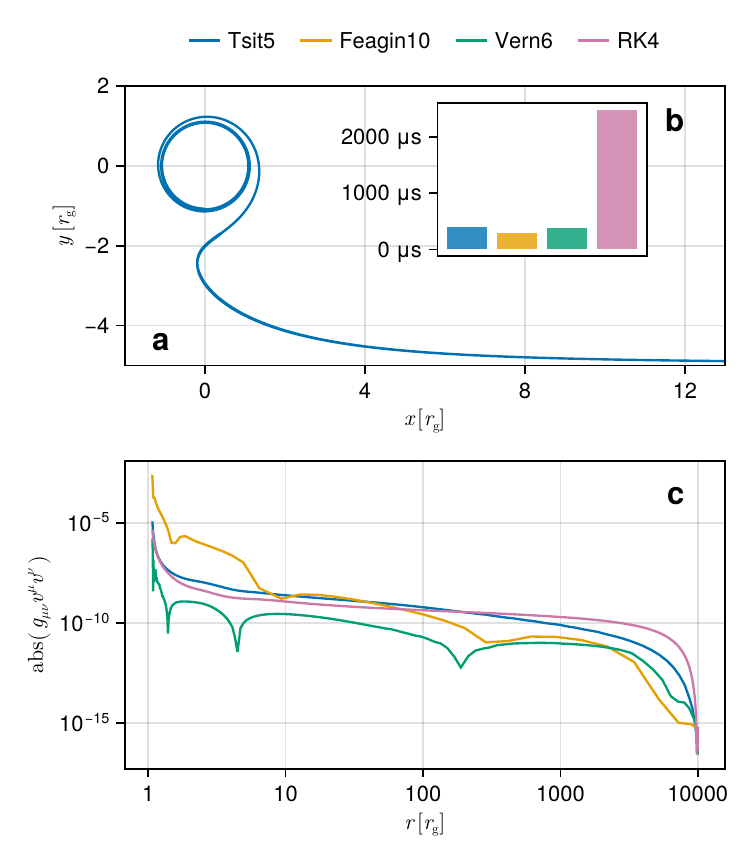}
    \caption{A comparison of different ODE integration algorithms using the
        default tolerances $\text{abstol} = \text{reltol} = 10^{-9}$. Panel a)
        shows the null-geodesic of the Kerr spacetime, $a = 0.998$, for impact
        parameters $\alpha = 5$ and $\beta = 0$ integrated from the starting
        position $x^r = 10^4 \rg$ and $x^\theta = \pi / 2$. All of the
        algorithms considered yield almost identical solutions. Panel b) shows
        the integration time for the geodesic considered. The integration time
        is approximately equal for all of the algorithms except for RK4, which
        is significantly longer. Panel c) shows the value of the conserved
        quantity $g_{\mu \nu} v^\mu v^\nu = 0$ along the trajectory, used as a
        measure of stability of the integration algorithm.
    }
    \label{fig:dot-stability}
\end{figure}

Our integration method does not use the geodesic constants of motion directly,
and so the stability of the integrator may be evaluated by calculating these or
other invariants along the trajectory. In Figure~\ref{fig:dot-stability} we show
the invariant $v_\mu v^\mu$ for a geodesic that spirals into a maximally
spinning black hole. The magnitude of the invariant is shown for a sample of
integration algorithms, along with the time-to-solution for the geodesic. Note
that this solving time includes the time to initialize the integrator, calculate
the full trajectory (with interpolants), and package the solution structure in
memory.

To assert the accuracy we calculate the angular deflection problem. The
deflection is the difference in $x^\phi$ of a geodesic traveling from positive
to negative infinity, that is
\begin{equation}
    \delta x^\phi :=
        \left.x^\phi\right\rvert_{+\infty} - \left.x^\phi\right\rvert_{-\infty}
        - \pi,
\end{equation}
where $\pi$ is the angular change for a trajectory that experiences no
deflection. Semi-analytic solutions for the deflection angle in the Kerr
spacetime have been calculated for equatorial geodesics in
\citet{iyer_lights_2009}. The authors use elliptical integrals to find the
coordinate differences. We follow their notation and denote the analytic
deflection angle $\hat{\alpha}$.

\begin{figure}
    \centering
    \includegraphics[width=0.94\columnwidth]{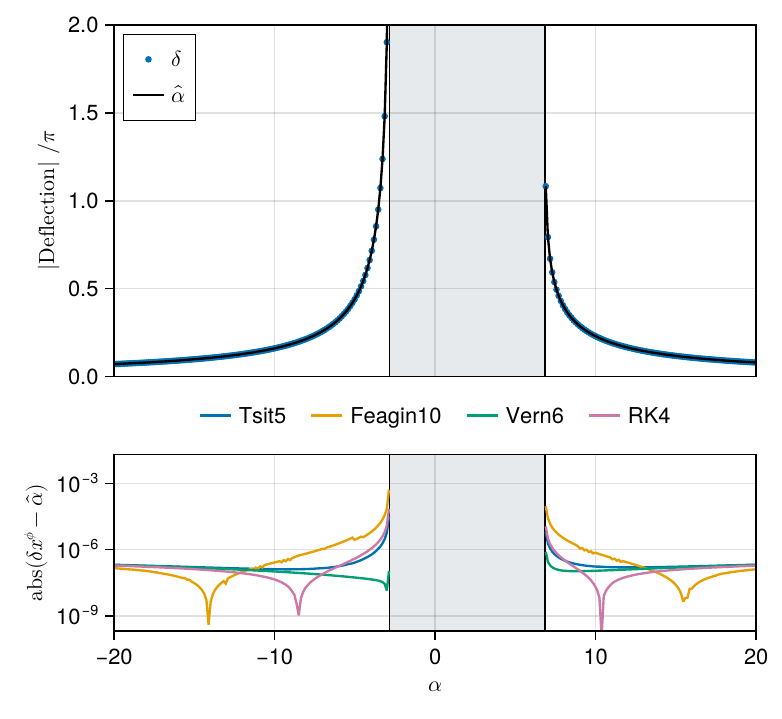}
    \caption{Deflection angle in the Kerr spacetime ($a = 0.998$) for geodesics
        in the equatorial plane over a range of impact parameters $\alpha$.
        Upper panel: numerical deflection $\delta x^\phi$ calculated with
        $x^r_\text{start} = 2 \times 10^8 \, \rg$, absolute and relative
        tolerances set to $10^{-14}$, and effective infinity $4 \times 10^8\,
        \rg$, shown with the numerical solutions for $\hat{\alpha}$. Lower
        panel: the absolute relative error between the numeric and analytic
        deflection angles for different integration algorithms.}
    \label{fig:deflection-angle}
\end{figure}

Figure~\ref{fig:deflection-angle} shows the deflection angle as a function of
impact parameter $\alpha$, along with a  measure of the error for the different
integration algorithms. There is asymptotic behaviour of the error as $\lvert
\alpha \rvert$ increases. This is related to our approximation of an `observer
at infinity' discussed in Section~\ref{sec:observers-and-emitters}. As can be
expected, the error increases if $x^r_\text{start}$ is decreased, and
vice-versa. Here we again see the impact that the choice of integrator can have
on numerical errors.

\subsection{Emissivity profiles}

\begin{figure}
    \centering
    \includegraphics[width=0.99\columnwidth]{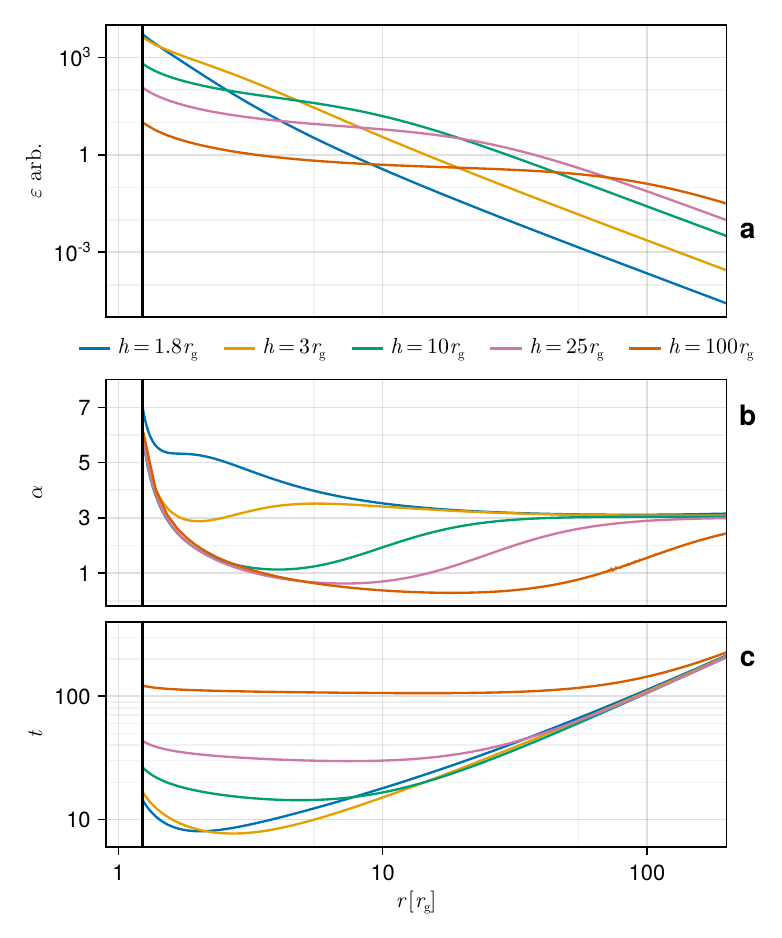}
    \caption{Emissivity profiles, $\varepsilon(\rho)$ for a lamppost point
        source illuminating a thin, equatorial disc at various heights above the
        spin axis of a maximally spinning Kerr black hole ($a = 0.998$). Panel
        a) shows the emissivity profiles in arbitrary flux units.  Panel b) is
        the $\alpha$ exponent of the emissivity profile found by differentiating
        the emissivity profile assuming $r^{-\alpha}$. Panel a) and b) are to be
        compared to Figures 2 and 3 in \citet{dauser_irradiation_2013}. Panel c)
        is the light-travel time of the photon from the lamppost to the disc.
        The increase in light travel time at small radii is due to the strong
        gravity effects.
}
    \label{fig:emissivity-profiles}
\end{figure}

To test our implementation of the emissivity profile calculations described in
Section~\ref{sec:emissivity-profiles}, we model a thin equatorial accretion disc
illuminated by an on-axis lamppost corona, comparing against
\citet{wilkins_understanding_2012} and \citet{dauser_irradiation_2013}. These are
shown in Figure~\ref{fig:emissivity-profiles} and are in good agreement with the
published results. Also shown is the effect of the Shapiro delay $t$, on the
arrival time of a photon at a given radius on the disc.

\subsection{Line profiles}

\begin{figure}
    \centering
    \includegraphics[width=0.99\columnwidth]{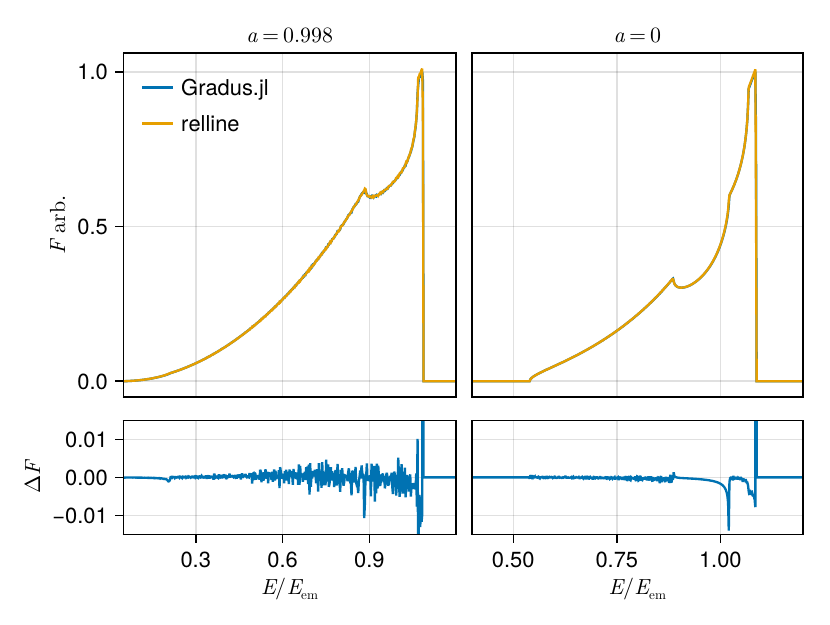}
    \caption{Comparison of line profiles calculated by integrating transfer
        functions with emissivity $I_\text{em} = \varepsilon(\rhoem) =
        \rhoem^{-3}$ using \Gradus and \relline. The transfer functions are
        calculated for an observer at $r_\text{obs} = 1000\rg$ and
        $\theta_\text{obs} = 40^\circ$, and integrated between $r_\text{in} =
        \risco$ and $r_\text{out} = 50 \rg$. Left panel is the maximally
        spinning Kerr spacetime, whereas the right panel is the Schwarzschild
        spacetime.}
    \label{fig:relline-comparison}
\end{figure}

The rest-frame accretion disc spectrum is modified by relativistic effects
producing an observed spectrum in which emission lines have a broad and skewed
\emph{line profile}.  To calculate these line profiles, transfer functions are
integrated for a particular emissivity profile as described in
Section~\ref{sec:transfer-function-integration}, neglecting the timing
components. Figure~\ref{fig:relline-comparison} compares the line profiles
computed using the transfer functions of \Gradus and the \relline model of
\citet{dauser_broad_2010}\footnote{We compare against the \relline v2.3 with
    table v0.5a distributed in the Relxill package
    \url{http://www.sternwarte.uni-erlangen.de/~dauser/research/relxill/}. These
are the latest versions at the time of writing.}. The spikes in residual
coincide with discontinuous features in the lineprofiles, and stem from minor
differences in the estimates of the extremal $g$. Very small differences here
can shift the edges of the lineprofile, resulting in residual spikes that can be
largely ignored. Overall, we see good agreement to within $\sim 1\%$ relative
difference across all parameters.

\subsection{Reverberation lags}
\label{sec:lag-transfer-functions}

The time delay between the corona's direct and the subsequent disc's reflected
components is increased by the effects of strong gravity on the light travel
time. The effect, termed \emph{reverberation lag}, is therefore dependent on
properties of the black hole, as well as depending on the disc and corona
geometry \citep[see e.g.][for review]{uttley_x-ray_2014,
cackett_reverberation_2021}

\begin{figure}
    \centering
    \includegraphics[width=0.97\columnwidth]{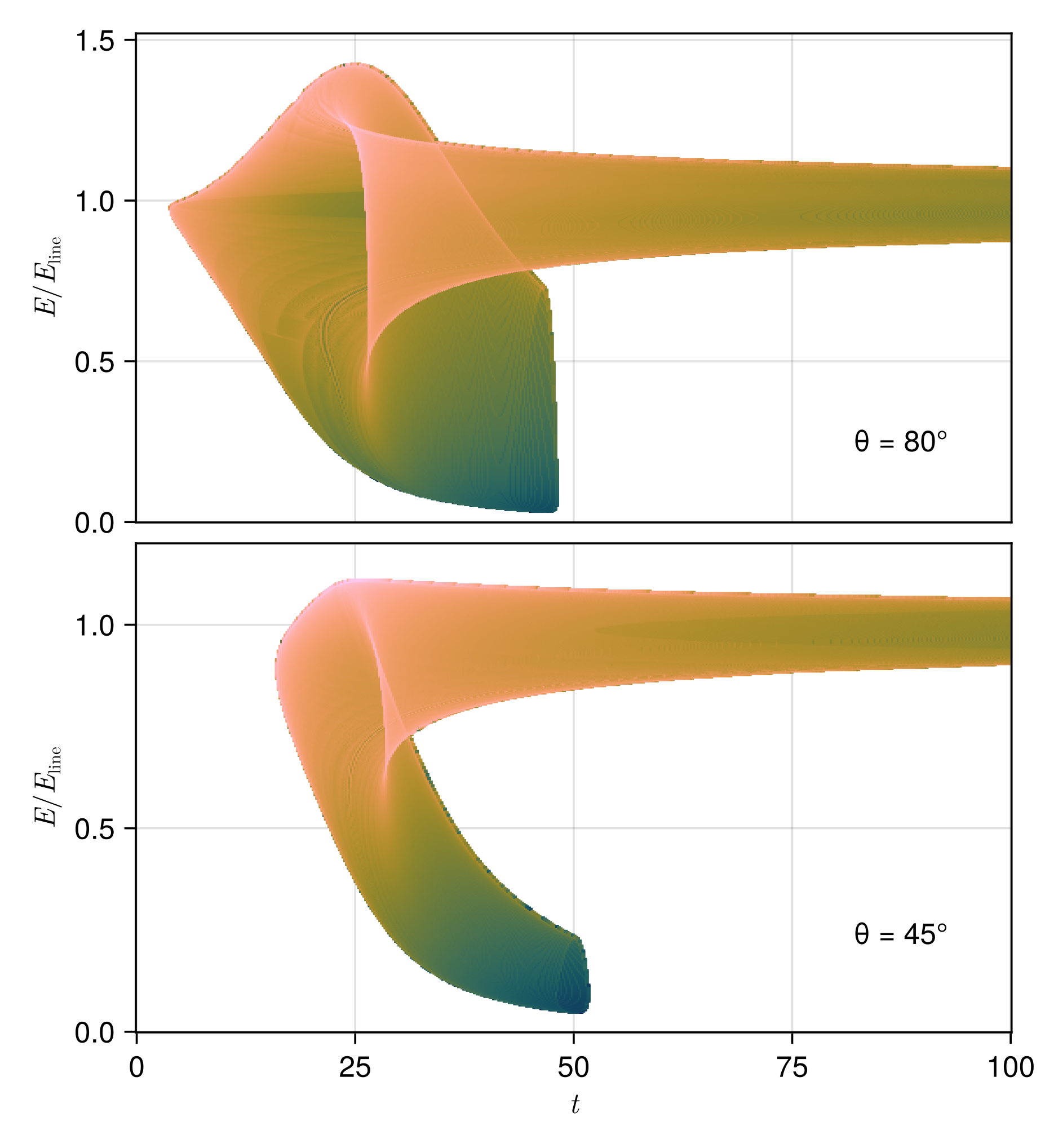}
    \caption{Two-dimensional transfer functions for the maximally spinning Kerr
        spacetime ($a = 0.998$) for two different observer inclinations. The
        colouring represents the logarithm of the flux in that particular bin in
        arbitrary units. The brighter the bin, the more flux is present. These
        are calculated by integrating the time-dependent Cunningham transfer
        functions
    (details in the text).}
    \label{fig:lag-frequency-transfer-functions}
\end{figure}

Simulating either the lag--frequency or lag--energy spectra involves computing a
set of transfer functions that record the arrival time and observed energy of
each flux element \citep{reynolds_x-ray_1999}. We distinguish these transfer
functions from others by referring to them as the \textit{lag transfer
functions}. Two examples are shown in
Figure~\ref{fig:lag-frequency-transfer-functions}. By convention the origin of
the time axis is set to the arrival time of the continuum emission and is
unambiguous for a lamppost corona. We here test our method for calculating high
resolution lag transfer functions using the Cunningham transfer functions,
described in Section~\ref{sec:transfer-function-integration}.

% The lag transfer functions depend on the choice of observer inclination,
% spacetime, coronal model, and disc model. The inclination of the observer is
% sensitive to gravitational lensing effects as well as influencing the relative
% distances of different regions of the disc and corona, affecting the arrival
% times of emission. The spacetime changes both the light travel time close to the
% central singularity and the gravitational redshift. The spacetime is therefore
% imprinted in both the energy and arrival time of each flux element.  The coronal
% model, in its geometry and spectra, modifies the emissivity profile of the
% accretion disc (which in turns changes the flux element coming from each disc
% element). The coronal geometry may also change the arrival time of both the
% continuum and reflected emissions. The disc geometry contributes to all of the
% above.

% \citep{reynolds_x-ray_1999,wilkins_origin_2013,cackett_modelling_2014}

\subsubsection{Lag--frequency spectra}

Summing the lag transfer functions
(Figure~\ref{fig:lag-frequency-transfer-functions}) over the energy axis for a
given energy range yields an \textit{impulse response function} $\psi(t)$, which
encodes how the disc responds to the impulse of a coronal flash.

\begin{figure}
    \centering
    \includegraphics[width=0.98\columnwidth]{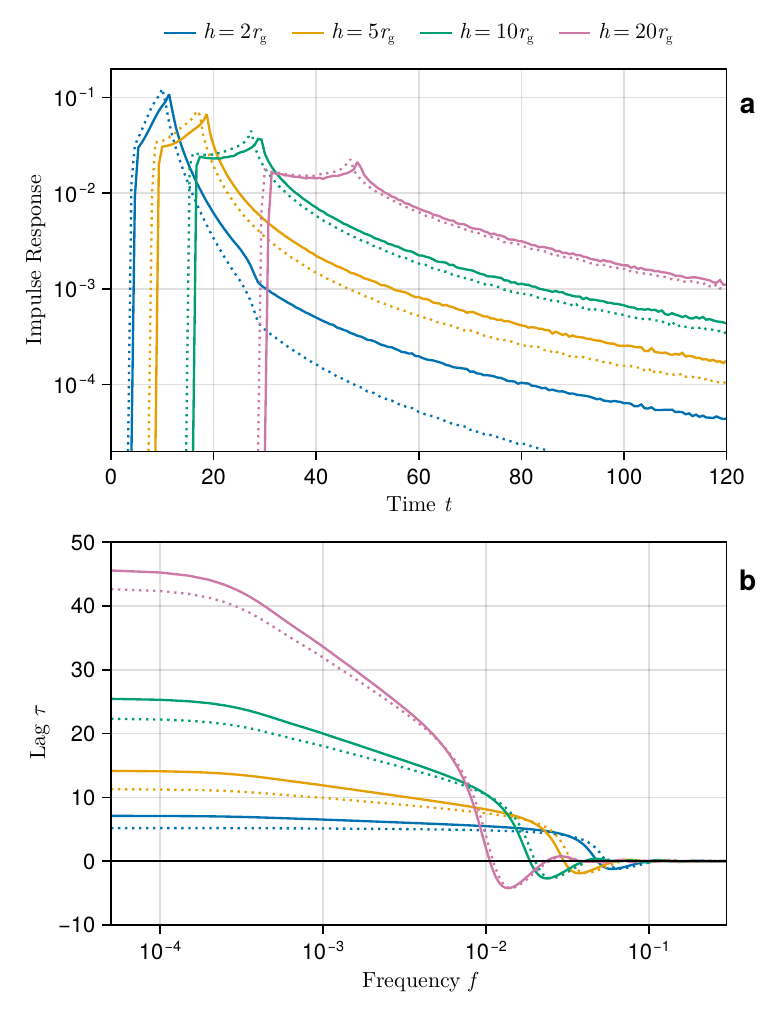}
    \caption{Impulse responses and lag--frequency spectra of a lamppost model in
        the Kerr spacetime ($a = 0.998$) for different heights of the lamppost
        model. The solid lines are the razor-thin disc case, whereas the dotted
        lines are for the Shakura--Sunyaev disc solution with $\dot{M} /
        \dot{M}_\text{Edd} = 0.3$. Panel a) shows the impulse responses summed
        across all energy bands, and panel b) shows the corresponding lag--frequency
        spectra of the impulse responses. }
    \label{fig:reverberation-thin}
\end{figure}

Examples for different lamppost heights over the full energy range are shown in
the top panel of Figure~\ref{fig:reverberation-thin}.

Following \citet{cackett_modelling_2014} (hereafter C14), we define the
\textit{response fraction}, $R$, as the ratio of reflected to continuum flux.
The impulse response in the Fourier domain is the scaled Fourier transform
\begin{equation}
    \mathscr{F}_\psi(f) := R \int_{0}^\infty \psi(t) \e^{-2\pi i f t} \d t
\end{equation}
at frequency $f$.
The phase difference between the reflected and continuum flux is
\begin{equation}
    \phi(f) = \tan^{-1} \left(
        \frac{\Im{\mathscr{F}_\psi}}{1 + \Re{\mathscr{F}_\psi}}
    \right),
\end{equation}
where $\Im{\mathscr{F}_\psi}$ and $\Re{\mathscr{F}_\psi}$ are the imaginary and
real components of $\mathscr{F}_\psi$ respectively. The imaginary component
represents the lag contribution to the phase difference. Since the driving
signal is present in both bands, it adds no lag contribution, but serves to
dilute the phase difference and therefore the real component of the signal
through the $+1$ in the denominator.

A subtlety to address here in the normalisation of the impulse responses, as we
do not fully model the continuum spectrum. We assume, as in
C14, that the reflected flux of the line is equal to
the continuum flux ($R = 1$). We normalise the impulse response functions by
dividing by a factor $Q = \int \psi_{\text{Fe K}\alpha}(t)$, so that the area
under the line impulse response function is unity.

The time lag is defined as
\begin{equation}
    \tau(f) := \frac{\phi}{2 \pi f},
\end{equation}
and relates the observed time lag to the Fourier frequency of the driving
signal (lower panel of Figure~\ref{fig:reverberation-thin}).

We find some small differences with C14. Our impulse responses and lags are in
good agreement for low coronal height, but start to differ by a few percent as
the corona is moved to $h \geq 10 \rg$. The impulse response in C14 starts
later, and the low frequency lag is longer with increasing $h$. This difference
is consistent with a constant multiplicative factor of $\sim 1.1$ in the time
axis of the impulse responses. The precise shapes of our impulse responses
agree, merely their offset differs. These differences are consistent with a
systematic offset in either the corona-to-disc, corona-to-observer, or
disc-to-observer time, or some combination thereof. Most likely it is some
disagreement in the corona-to-disc time, but with much modification to our code
we are unable to reproduce exactly the results of C14. Our light travel times
are consistent with the results of e.g. \citet{wilkins_origin_2013}, and we
believe the systematic difference is not in our implementation.

\subsubsection{Lag-energy spectra}

\begin{figure}
    \centering
    \includegraphics[width=0.98\columnwidth]{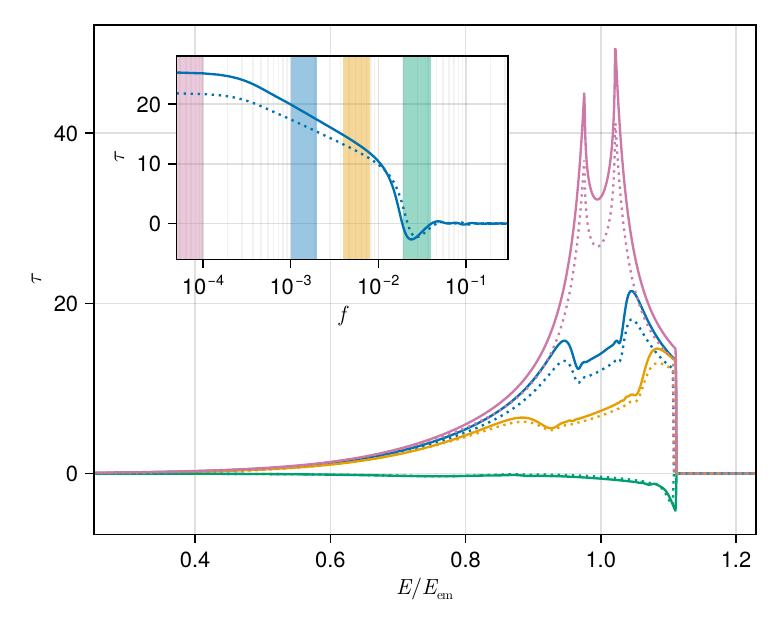}
    \caption{Lag--energy spectra for the same setup as in
        Figure~\ref{fig:reverberation-thin} but only for case where the lamppost
        height is $h=10$. The different colours now correspond to the frequency
        bands used to calculate the lag--energy spectrum (shown in the inset
        panel). The solid lines, as before, denote the razor thin disc, whereas
    the dotted lines are the Shakura--Sunyaev disc with $\dot{M} /
\dot{M}_\text{Edd} =0.3$.}
    \label{fig:lag-energy}
\end{figure}

Lag--energy spectra show the time lag as a function of energy within specific
frequency bands $f + \Delta f$. Since we have mandated for consistency with C14
that the reflected flux of the \FeKa line is equal to the continuum flux, the
reference energy band is $E/E_\text{em} = 1$.  For the impulse response of each
energy channel (rows in the lag transfer functions), the lag--frequency spectrum
is calculated. The mean lag within $f + \Delta f$ is determined, and plotted as
a function of energy. Figure~\ref{fig:lag-energy} shows the time lag as a
function of energy relative to the \FeKa band, to be compared to Figure 11 in
C14.

\subsection{Radiative transfer}

\begin{figure*}
    \centering
    \includegraphics[width=0.99\linewidth]{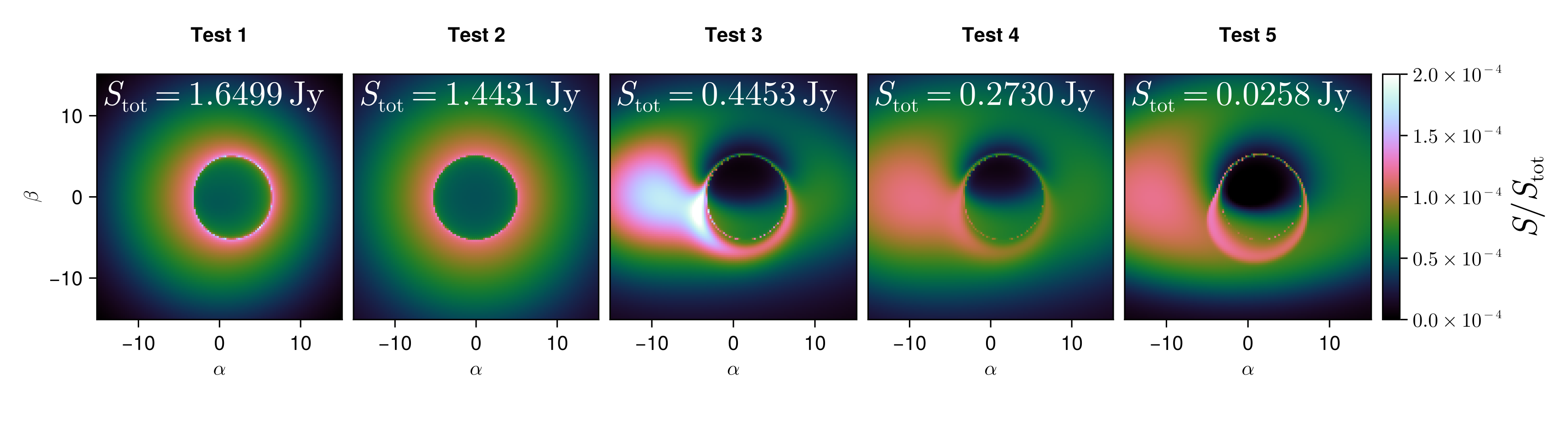}
    \caption{Intensity images calculated with \Gradus of the radiative transfer
    analytic test models specified in \citet{gold_verification_2020} with
    resolution $128 \times 128$ pixels, and impact parameters ranging between $-15\,
    \rg$ and $15\, \rg$, and observer position $r_\text{obs} = 1000\, \rg$ and
    inclination $\theta_\text{obs} = 60^\circ$. The test cases correspond to the
    test parameters in their Table 1. The colouring is the intensity for the
    geodesic corresponding to that pixel normalized over the total intensity .}
    \label{fig:gold-test-problems}
\end{figure*}

\citet{gold_verification_2020} specify an analytic model for testing radiative
transfer codes (their Section 3.2, with results shown in their Figure 2 and 3).
The model gives the emissivity and absorptivity coefficients of a (corotating)
fluid as a function of radial coordinate. There are five free parameters in this
specification that can be used to control the model, for which they give 5
standardized tests.

We have implemented their test cases and see very good agreement, shown in
Figure~\ref{fig:gold-test-problems}.

\subsection{Other spacetimes}

\begin{figure}
    \centering
    \includegraphics[width=0.99\columnwidth]{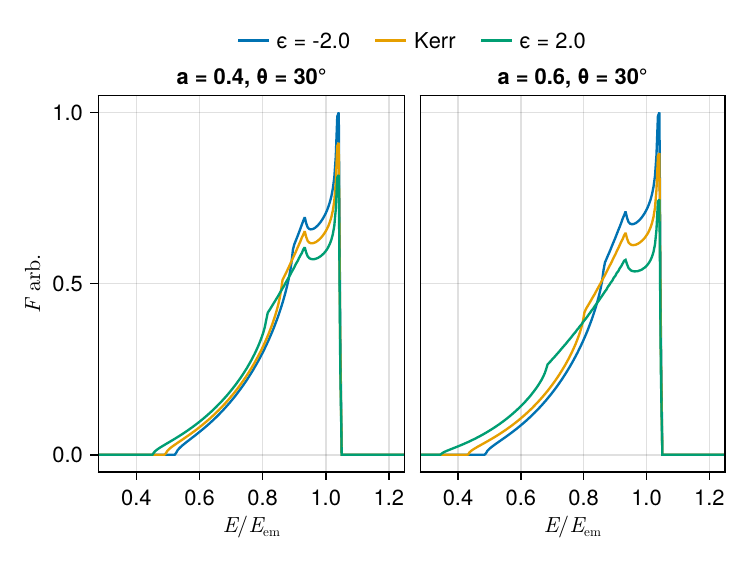}
    \caption{Sample lineprofiles for the various values of the deformation
    parameter $\epsilon$ for the metric described in
    \citet{johannsen_testing_2013}. Both cases assume a powerlaw emissivity profile
    with emissivity profile $\rhoem^{-3}$.}
    \label{fig:reflection-johannsen}
\end{figure}

\begin{figure}
    \centering
    \includegraphics[width=0.99\columnwidth]{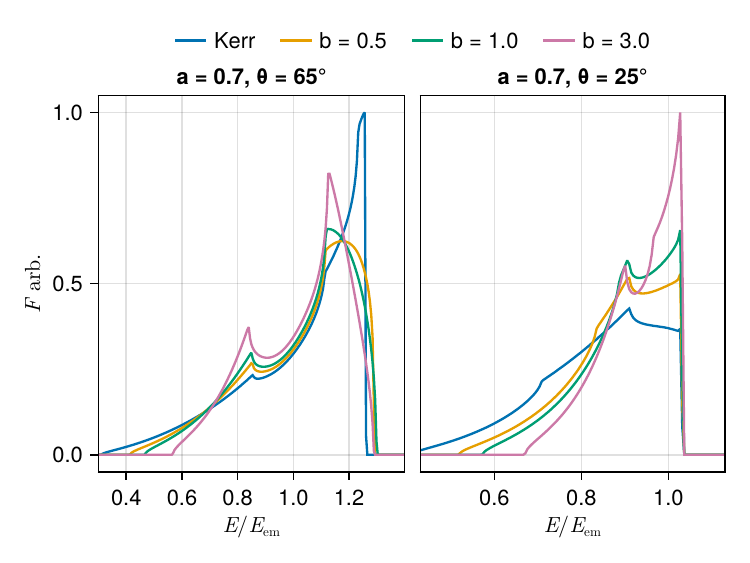}
    \caption{The same as in Figure~\ref{fig:reflection-johannsen} but for the
    metric described in \citet{garcia_class_1995}.}
    \label{fig:reflection-emda}
\end{figure}

To test our implementation for other spacetimes, we compare the line
profiles calculated with \Gradus to those calculated in
\citet{johannsen_testing_2013}. Our lineprofiles, calculated via Cunningham
transfer function integration, are shown in
Figure~\ref{fig:reflection-johannsen}. We find excellent agreement with the
published line profile. We demonstrate the flexibility of the code by
implementing the Einstein--Maxwell--Dilaton--Axion metric derived in
\citet{garcia_class_1995} and calculate lineprofiles for various values of the
$b$ constant, shown in Figure~\ref{fig:reflection-emda}. These Cunningham
transfer functions may be tabulated and readily used in spectral models.

\section{Applications to thick discs}
\label{sec:applications}

The Cunningham transfer function algorithm described in
Section~\ref{sec:partially-obscured-functions} is here used to calculate line
profiles and reverberation lag features from a lamppost corona for the SSD,
similar to, and in agreement with work by other authors
\citep{taylor_exploring_2018,taylor_x-ray_2018}.

\begin{figure}
    \centering
    \includegraphics[width=0.99\columnwidth]{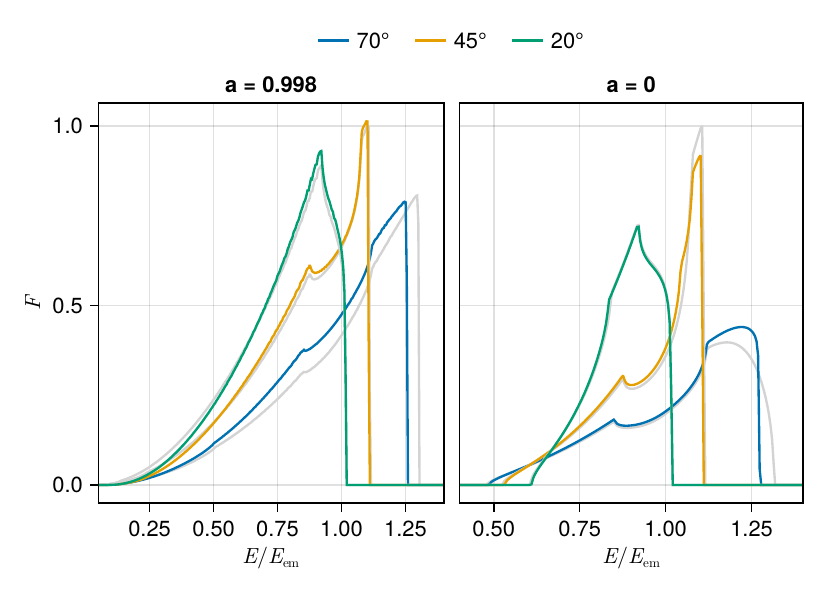}
    \caption{Line profiles for the SSD with $\dot{M} / \dot{M}_\text{Edd} = 0.3$
    for different observer inclinations $\theta_\text{obs}$ and emissivity
$I_\text{em} = \varepsilon(\rhoem) = \rhoem^{-3}$. The transfer functions are
calculated as in Figure~\ref{fig:relline-comparison}, and integrated over the
same limits. The light-grey lines correspond to the geometric thin disc
($\dot{M} / \dot{M}_\text{Edd} = 0$), and differ only for steep inclinations due
to obscuration of the inner $\rhoem$. Left panel is the the maximally spinning
Kerr spacetime, whereas the right panel is the Schwarzschild spacetime.}
    \label{fig:line-profile-ssd}
\end{figure}

Illustrative results for the line profiles with a fixed emissivity function
$\varepsilon(\rhoem) = \rhoem^{-3}$ for the SSD are shown in
Figure~\ref{fig:line-profile-ssd}. The line profiles are similar to the razor
thin disc case, except at steep inclination, where the blue shifted
contributions of the disc are reduced. As we can approximately consider the
different energy ranges to originate from different radii on the disc
\citep{gates_on_2024}, we can interpret the thick disc as obscuring the blue
shifted contributions. This obscuration is preferential due to the rotation of
the black hole generating an asymmetry in the blue and red shifted sides of the
accretion disc, see Figure~
\ref{fig:transfer-parameterisation}.

\begin{figure*}
    \centering
    \includegraphics[width=0.99\linewidth]{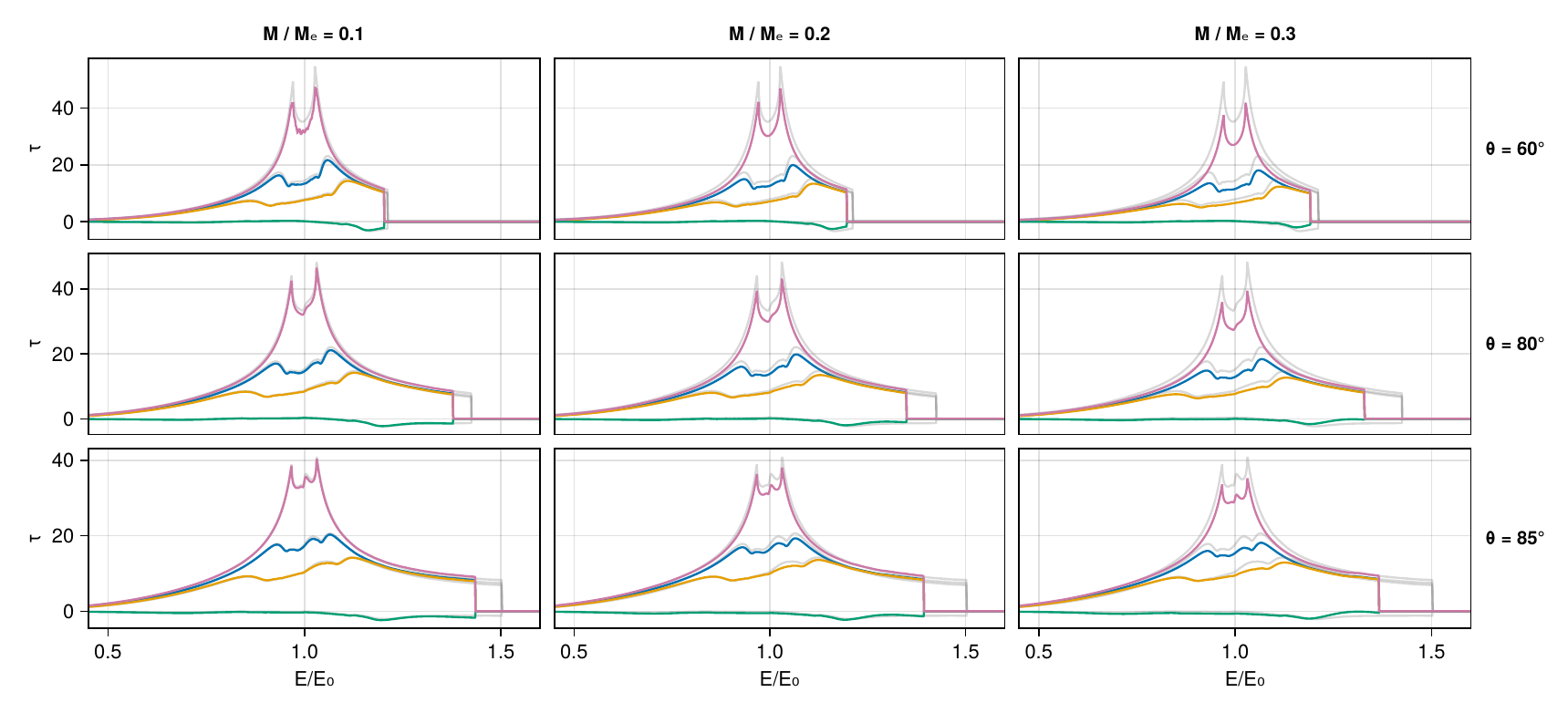}
    \caption{Lag--energy profiles for the SSD, using the same colour scheme as in
        Figure~\ref{fig:lag-energy}. For all figures the Kerr spacetime
        ($a=0.998$) is used with a lamppost corona, $h = 10 \rg$. The light grey
        lines show the corresponding razor-thin disc lag--energy profiles. The
        columns show the effect of changing the Eddington ratio $\dot{M} /
        \dot{M}_\text{Edd}$, and the rows are changing the inclination. When
        $\theta \lesssim 40^\circ$, the differences between the thin disc and
        the SSD are minimal when the lamppost corona is at an appreciable height
        above the black hole.}
    \label{fig:reverb-thick-discs}
\end{figure*}

Our reverberation lag results, for a fixed corona height $h=10 \rg$ with
different Eddington ratios, are shown in Figure~\ref{fig:reverb-thick-discs}. We
interpret these results similar to the impact on the lineprofiles, in that only
at steep inclinations, where obscuration is prominent, is a pronounced
difference between the thin and thick disc case visible. There is also generally
a decrease in the lag across the energy range due to the reduced distance
photons must travel to hit the surface of the disc.

At lower inclinations, below $60^\circ$, the effect of the thick disc in the
lag--energy spectrum for a corona at $h \geq 10 \rg$ is negligible, even at
moderate Eddington ratios $\dot{M} / \dot{M}_\text{Edd} = 0.3$. From an
observational perspective, the reduced lag for such a configuration would likely
only be detectable in the high energy deficit, however such a detection would
require a steep inclination angle, likely accompanying high degrees of
extinction and obscuration from other regions in the plane of the disc or torus.

\begin{figure*}
    \centering
    \includegraphics[width=0.99\linewidth]{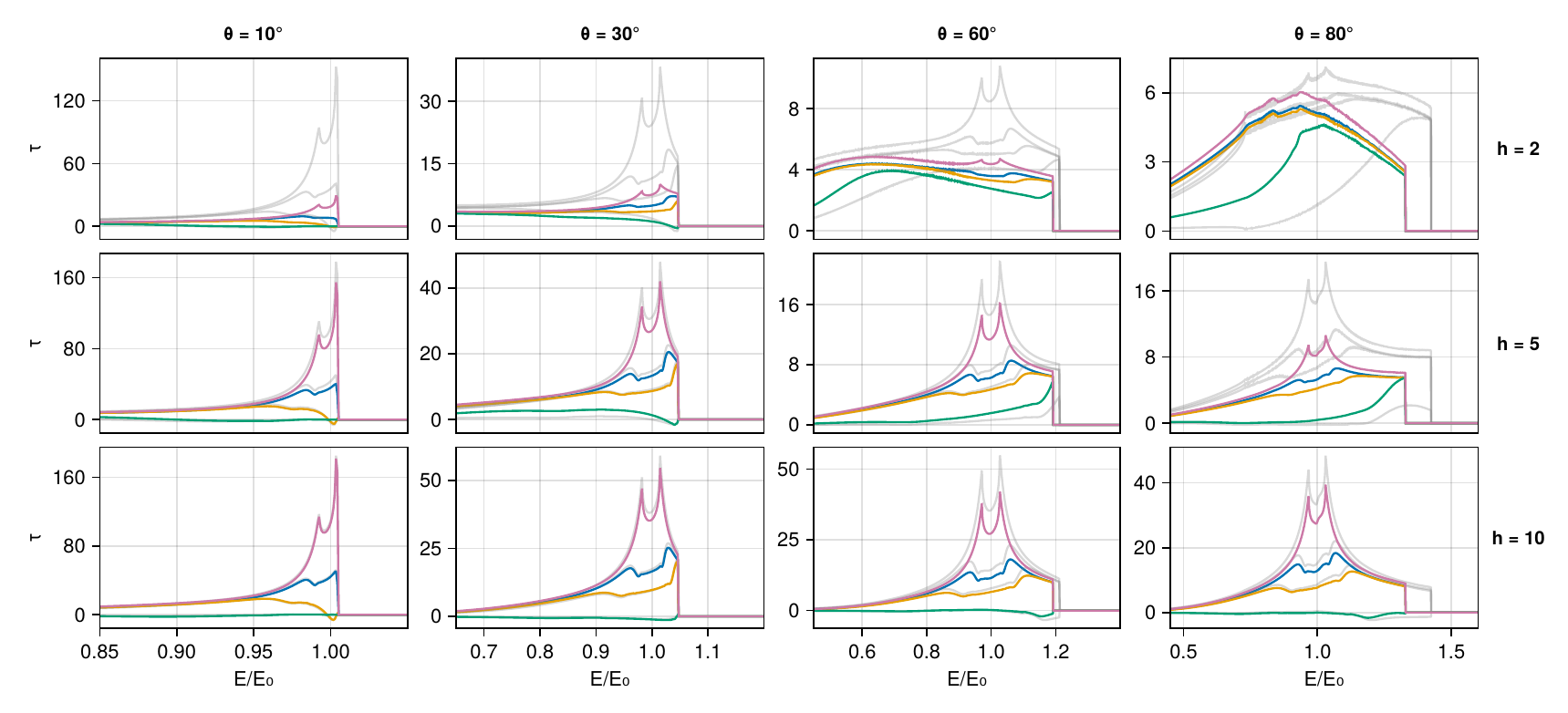}
    \caption{As in Figure~\ref{fig:reverb-thick-discs}, except the Eddington
        ratio is now fixed to $0.3$, and instead the lamppost corona height is
        varied. When the height of the corona is low, there is significant
        obscuration leading to increased emissivity at low radii, and reduced
        emissivity at distant radii, and similarly with the light crossing
    times. This makes a marked change irrespective of inclination, but only
occurs when the coronal height is comparable to the height of the disc.}
    \label{fig:reverb-thick-discs-corona}
\end{figure*}

In Figure~\ref{fig:reverb-thick-discs-corona}, the lag--energy spectrum for
different lamppost heights is shown. A low corona has a much more prominent
effect than a high corona, greatly reducing the lag and changing the shape of
the spectrum. These effects can be principally attributed to two reasons. First,
the much greater relative difference in light travel time from the corona to the
thick disc when the lamppost is low, compounded by the light--crossing times due
to the proximity to the black hole. This means the reprocessed reflected
component occurs much earlier, diluting the lag significantly across the entire
energy range. Second, the reduced illumination at distant radii on the disc
suppresses the emissivity profile and therefore flux contributions from those
regions. This will reduce the impulse response around the line energy $E /
E_\text{em} =
1$, in turn diluting the lag at difference Fourier frequencies, and changing the
shape of the lag--enregy spectra.

Notably, for a low corona, a marked difference between the thin and thick disc
lag--frequency and lag--energy profiles exists for all inclinations. The shift
in the lag is not precisely equivalent to a change in the coronal height, as the
emissivity profiles of a geometrically thick disc are modified from the
razor-thin counterparts \citep{taylor_x-ray_2018}. With the current resolution
and signal to noise of contemporary instruments, it would be difficult to
disambiguate the effect of the disc height and coronal height when $h \sim 3 - 10
r_\text{g}$. The shape of the lag--energy spectrum for these heights for the thin
and thick discs is remarkably similar, however reduced in the case of the
latter. Especially at lower inclination, if the disc were thick but the model
used to fit the data assumes a thin geometry, the recovered lag would be an
overestimate, with slight ramifications on the black hole mass deduced from the
lag time.

\section{Conclusions}
\label{sec:conclusion}

\Gradus is a new open-source and publicly available general relativistic
ray-tracing toolkit. It is designed to be extensible to a wide number of
problems, and uses sophisticated algorithms to produce high-resolution
simulations efficiently. The numerical methods and some implementation details
have been discussed, and \Gradus has been tested against a number of standard
problems in the literature. \Gradus has applied to calculating line profile and
reverberation results for the standard Shakura--Sunyaev accretion disc, and the
effects of obscuration and lamppost corona height discussed. Illustrative line
profile simulations for non-Kerr space times have also been shown, and any
result calculated with Gradus can switch the spacetime trivially, expediting
the process of developing new codes for new metrics.

Our software can reliably compute the Cunningham transfer functions used in
spectral models for a variety of accretion flows. This permits us to create fast
line profile models that treat e.g. both the lamppost height and Eddington ratio
of a SSD as free parameters, and allowing us to build on the work of e.g.
\citet{jiang_black_20222}. Anticipated applications are to calculate transfer
function tables for turbulent velocity profiles \citep{pariev_line_1998}, for
warped accretion discs \citep[e.g.][]{hartnoll_reprocessed_2001}, for conical or
sub-Keplerian discs \citep[e.g.][]{wu_iron_2007}, and more.

We are happy to assist with new applications of \Gradus, and encourage members
of the community to contact us with interesting problems. Our ultimate aim is to
make \Gradus a useful tool for quickly exploring and producing X-ray spectral
models of compact objects.

\section*{Acknowledgements}
This work was supported by the UKRI AIMLAC CDT funded by grant EP/S023992/1. This
work was supported by the Science and Technology Facilities Council grant number
ST/Y001990/1.

We thank the reviewer for their helpful and constructive comments. We thank
Jiachen Jiang, Cosimo Bambi and Askar Abdikamalov for sharing their software for
comparisons, and Corbin Taylor for making the \software{fenrir} code, along with
many example scripts, public at our request. FB thanks Rosaline von Hardturm for
her expert debugging assistance. All figures created using Makie.jl
\citep{DanischKrumbiegel2021}, using the color scheme of
\citet{wong_points_2011}.

%%%%%%%%%%%%%%%%%%%%%%%%%%%%%%%%%%%%%%%%%%%%%%%%%%
\section*{Data Availability}

No new data or analyses have been created for this work. The code to reproduce
this paper and all figures therein is freely available under GPL 3.0 license:
\url{https://github.com/fjebaker/gradus-paper}

% The inclusion of a Data Availability Statement is a requirement for articles published in MNRAS. Data Availability Statements provide a standardised format for readers to understand the availability of data underlying the research results described in the article. The statement may refer to original data generated in the course of the study or to third-party data analysed in the article. The statement should describe and provide means of access, where possible, by linking to the data or providing the required accession numbers for the relevant databases or DOIs.



\bibliographystyle{mnras}
\bibliography{citations} % if your bibtex file is called example.bib

%%%%%%%%%%%%%%%%%%%%%%%%%%%%%%%%%%%%%%%%%%%%%%%%%%

%%%%%%%%%%%%%%%%% APPENDICES %%%%%%%%%%%%%%%%%%%%%

\appendix

\section{Orthonormalization with Gram--Schmidt}
\label{appendix:gram-schmidt}

The theorem of Gram--Schmidt states that it is always possible to construct a set of orthonormal vectors in any inner-product space $\mathbb{R}^n$, and uses a projection-subtraction procedure as a proof \citep{schmidt_uber_1989}. Starting with $n$ linearly independent vectors $\vector{v}_n$, and denoting the projection of a vector $\vector{u}$ along the direction of $\vector{v}$ as
\begin{equation}
\mathrm{P}_{\vector{v}}\left(\vector{u}\right) := \frac{\vector{v} \cdot \vector{u}}{\vector{u} \cdot \vector{u}}\ \vector{u} = \frac{g_{\mu\nu} v^\mu u^\nu}{g_{\sigma\rho} u^\sigma u^\rho} \vector{u},
\end{equation}
allows expressing the Gram-Schmidt procedure as
\begin{align}
    \vector{k}_1 &= \vector{v}_1, \nonumber \\
    \vector{k}_2 &= \vector{v}_2 - \mathrm{P}_{\vector{k}_1}\left(\vector{v}_2 \right), \nonumber \\
    &\vdots \nonumber \\
    \vector{k}_n &= \vector{v}_n - \sum_{i = 1}^{n-1} \mathrm{P}_{\vector{k}_i} \left(\vector{v}_n \right).
\end{align}
Constructing meaningful orthonormal frames requires appropriate choice of the initial linearly independent vectors $\vector{v}$, in order to associate global directions with the tetrad. The locally non-rotating frame (LNRF), with angular velocity $\omega = -g_{t\phi} / g_{\phi\phi}$, has tangential frame velocity $v^\mu = A (1, 0, 0, \omega)$ where $A$ is some normalization. To construct the LNRF, a choice of initial vectors may therefore be
\begin{align}
    \vector{v}_1 &= \left(1, 0, 0, \omega \right) \mapsto \dtensor{\e}{(t)}{\mu}, \nonumber \\
    \vector{v}_2 &= \left(1, 0, 0, 1\right) \mapsto \dtensor{\e}{(\phi)}{\mu}, \nonumber \\
    \vector{v}_3 &= \left(1, 1, 0, 1\right) \mapsto \dtensor{\e}{(r)}{\mu}, \nonumber \\
    \vector{v}_4 &= \left(1, 1, 1, 1\right) \mapsto \dtensor{\e}{(\theta)}{\mu},
\end{align}
where we have denoted the corresponding tetrad vector generated by the orthonormalization procedure after the arrow.

Other sensible frames require different initial vectors, and care must be taken in implementing a method that correctly reorders the resulting tetrad vectors: for example, the zero angular momentum (ZAMO) frame for an on-axis coronal source with velocity $\dot{x}^\mu = (1, \d r / \d t, 0, 0)$ requires
\begin{align}
    \vector{v}_1 &= \left(1, \d r / \d t, 0, 0 \right) \mapsto \dtensor{\e}{(t)}{\mu}, \nonumber \\
    \vector{v}_2 &= \left(1, 1, 0, 0\right) \mapsto \dtensor{\e}{(r)}{\mu}, \nonumber \\
    \vector{v}_3 &= \left(1, 1, 1, 0\right) \mapsto \dtensor{\e}{(\theta)}{\mu}, \nonumber \\
    \vector{v}_4 &= \left(1, 1, 1, 1\right) \mapsto \dtensor{\e}{(\phi)}{\mu}.
\end{align}

Our implementation of the Gram--Schmidt procedure is accurate up to machine-level with the analytic tetrads for the LNRF in \cite{bardeen_rotating_1972}, their Equation (3.2), and for the moving source ZAMO frame in \cite{gonzalez_probing_2017}, their Equation (10).

\section{Keplerian orbits of stationary, axis-symmetric spacetimes with accelerated geodesics}
\label{appendix:circular-orbits}

Keplerian circular orbits are orbits in the equatorial plane with velocity of
the form $v^\mu = A(1, 0, 0, \Omega)$, where $A$ is some normalization that
ensures \eqref{eq:velocity_constraint}, and
\begin{equation}
    \label{eq:keplerian-angular-velocity}
    \Omega := v^\phi / v^t,
\end{equation}
is the Keplerian angular velocity. We are therefore restricting ourselves to
$\theta = \pi/2$, and $v^r = v^\theta = 0$, and require stability through the
stationary point condition
\begin{equation}
    \frac{\d v^r}{\d \lambda} = 0.
\end{equation}

To begin, we will consider unaccelerated ($a^\mu = 0$) geodesics, and follow
\cite{johannsen_regular_2013} in rewriting \eqref{eq:geodesic_equation} as
\begin{equation}
    \frac{\d}{\d \lambda}\left( g_{\mu\sigma}v^\mu \right) = \frac{1}{2} v^\alpha v^\beta \partial_\sigma g_{\alpha \beta},
\end{equation}
where we have used the expansion of the Christoffel symbols
\eqref{eq:christoffel} and applications of the chain rule to manipulate the
form of the equation. Examining the radial component ($\sigma = r$) for
stationary, axis-symmetric spacetimes, along with $v^r = 0$, one obtains
\begin{equation}
    \label{eq:expanded-geodesic-equation}
    0 =
    \partial_r g_{tt} (v^t)^2
    + 2\partial_r g_{t\phi} v^t v^\phi
    + \partial_r g_{\phi\phi} (v^\phi)^2.
\end{equation}
Using \eqref{eq:keplerian-angular-velocity} yields
\begin{equation}
    \label{eq:omega-expression}
    \Omega =
    \left( \partial_r g_{\phi\phi} \right)^{-1}\left( \partial_r g_{t\phi} \pm \sqrt{\left( \partial_r g_{t\phi} \right)^2 - \partial_r g_{tt} \partial_r g_{\phi\phi}} \right),
\end{equation}
now determined entirely by metric components. Using
\eqref{eq:velocity_constraint}, mandating $\mu \neq 0$, and the stationary
point conditions $(v^r)^2 = 0$ and $\partial_r (v^r)^2 = 0$, one may continue
to find expressions for $E = -v_t$ and $L_z = v_\phi$ entirely in terms of
$\Omega$, the metric, and $\mu$. The algebra involved is straightforward but
verbose, and in the interest of brevity we will only state the results:
\begin{align}
    \frac{E}{\mu} &= \pm \mathcal{A} \left(g_{tt} + g_{t\phi}\Omega\right) , \label{eq:energy-of-orbit} \\
    \frac{L_z}{\mu} &= \pm \mathcal{A} \left(g_{t\phi} + g_{\phi\phi}\Omega\right), \\
    \mathcal{A} &= \left(\sqrt{-g_{\phi\phi} \Omega^2 - 2g_{t\phi} \Omega - g_{tt}}\right)^{-1}.
\end{align}
The two solutions correspond to prograde or retrograde orbits. Since these
expressions arise only from the velocity invariance, they are dependent on
$a^\mu$ only insofar as that $\Omega$ is dependent on $a^\mu$.

For accelerated geodeiscs, the acceleration vector modifies
\eqref{eq:expanded-geodesic-equation} with the addition of a $g_{rr} a^r$ term,
under the same assumptions of vanishing radial and poloidal velocity. Solving
for $\Omega$ as in \eqref{eq:omega-expression} is then trivial only when $a^r$
contains quadratic terms of $v^t$ and $v^\phi$, otherwise factors of $v^t$ and
$v^\phi$ must be substituted with $v^t = \pm \mu \mathcal{A}$ and $v^\phi = \pm
\mu \mathcal{A} \Omega$, resulting in polynomials of higher degree.

We will motivate our study by henceforth considering acceleration due to the
Faraday tensor,
\begin{equation}
    F_{\mu\nu} := \partial_\mu A_\nu - \partial_\nu A_\mu,
\end{equation}
where the potential driving the acceleration is axis-symmetric, $A_\mu = (A_t,
0, 0, A_\phi)$. Deviations thereof may not in general permit Keplerian circular
orbits without additional assumptions. Such axis-symmetric cases are still
useful in studying a number of interesting problems, including Kerr-Newman
spacetimes \citep{hackmann_charged_2013}, or black holes immersed in external
magnetic fields \citep{tursunov_circular_2016}.

For these axis-symmetric potentials, the radial acceleration for a particle with charge $q$ is
\begin{equation}
    a^r = qF^r_\mu x^\mu = q\left( F^r_t x^t + F^r_\phi x^\phi\right),
\end{equation}
and therefore, \eqref{eq:expanded-geodesic-equation} inherits an additional term
\begin{equation}
    0 = \textrm{\eqref{eq:expanded-geodesic-equation}} \pm q \frac{g_{rr}}{\mu \mathcal{A}} \left( F^r_t + F^r_\phi \Omega \right),
\end{equation}
which is quartic in $\Omega$ and must be solved numerically.

% \section{Some extra material}

%%%%%%%%%%%%%%%%%%%%%%%%%%%%%%%%%%%%%%%%%%%%%%%%%%

% Don't change these lines
\bsp    % typesetting comment
\label{lastpage}
\end{document}